\documentclass[aps,prd,floats,floatfix,nofootinbib,superscriptaddress]{revtex4-1}
\usepackage{graphicx,amsmath,amssymb,amsfonts,hyperref,mathdots,caption,subcaption, setspace, wrapfig,slashed}
\usepackage[usenames,dvipsnames,svgnames,table]{xcolor}

\newcommand{\unit}[1]{\ensuremath{\;\mathrm{#1}}}
\newcommand{\mr}{\ensuremath{m_{\zeta^{0,r}}}}
\newcommand{\mi}{\ensuremath{m_{\zeta^{0,i}}}}
\newcommand{\met}{\ensuremath{\slashed{E}_T}}
\newcommand{\DmMAX}{\ensuremath{\Delta m_0^{\mathrm{MAX}}}}
\newcommand{\zo}{\zeta^{0,r}}
\newcommand{\zi}{\zeta^{0,i}}
\newcommand{\multiplet}{\Sigma}
\newcolumntype{L}[1]{>{\raggedright\let\newline\\\arraybackslash\hspace{0pt}}m{#1}}
\newcolumntype{C}[1]{>{\centering\let\newline\\\arraybackslash\hspace{0pt}}m{#1}}
\newcolumntype{R}[1]{>{\raggedleft\let\newline\\\arraybackslash\hspace{0pt}}m{#1}}
\begin{document}

\title{LHC constraints on large scalar multiplet models with a $Z_2$ symmetry}

\author{Heather E.~Logan}
\email{logan@physics.carleton.ca}
\affiliation{Ottawa-Carleton Institute for Physics, Carleton University, Ottawa, Ontario K1S 5B6, Canada}

\author{Travis A. Martin}
\email{tmartin@triumf.ca}
\affiliation{TRIUMF, 4004 Wesbrook Mall, Vancouver, BC V6T 2A3, Canada}

\author{Terry Pilkington}
\email{tpilking@physics.carleton.ca}
\affiliation{Ottawa-Carleton Institute for Physics, Carleton University, Ottawa, Ontario K1S 5B6, Canada}

\date{July 7, 2015}

\begin{abstract}
We study the LHC search constraints on models that extend the Standard Model (SM) with an inert, complex scalar electroweak multiplet, $\multiplet$, with isospin $T=5/2$ (sextet) or $T=7/2$ (octet) and identical hypercharge to the SM Higgs doublet. Imposing a global $Z_2$ symmetry under which $\multiplet \to -\multiplet$, the lightest member of $\multiplet$ is stable and we require that it be neutral ($\zo$) to avoid cosmological constraints from charged relics.  Pair production of scalars by electroweak interactions followed by cascade decays to $\zo$ through $W$ and $Z$ emission produces signatures similar to those of supersymmetric electroweak gauginos, and we constrain the models by recasting a collection of such searches made with data from the 8~TeV run of the LHC.
We find that there is no sensitivity from these searches to the compressed spectrum regime, in which the mass splittings between the lightest and heaviest states in $\multiplet$ are less than about 20~GeV. 
In the remaining parameter space, we find significant exclusions for $\mr \sim 80-180$~GeV in the sextet model, and $\mr \sim 80-120$~GeV in the octet model.
\end{abstract}

\maketitle

\section{Introduction}
\label{sec:introduction}

The discovery of a Standard Model (SM)-like Higgs boson at the CERN Large Hadron Collider (LHC)~\cite{Aad:2012tfa,Chatrchyan:2012ufa} provides proof of the existence of weakly coupled scalar particles in nature.  Extensions of the scalar sector of the SM are thus of great interest, particularly when they could be connected to the solution of other mysteries of the SM, such as the nature of dark matter (DM) in the Universe~\cite{Gelmini:2015zpa}.

Extensions of the SM Higgs sector involving an additional ``inert'' scalar multiplet, the lightest state of which is stable and hence a possible dark matter candidate, have been well-studied in the singlet~\cite{Silveira:1985rk}, doublet~\cite{Deshpande:1977rw}, and triplet~\cite{Araki:2011hm} cases (for recent summaries of the experimental status of these models, see, e.g., Refs.~\cite{Cline:2013gha}, \cite{Belanger:2015kga} and~\cite{Ayazi:2014tha}). More recently, multiplets from larger representations of $SU(2)_L$ have been investigated in the context of dark matter~\cite{LargeMultipletDM,Earl:2013jsa,Earl:2013fpa}.  

In this paper, we expand upon the study in Ref.~\cite{Earl:2013fpa}, which focuses on models in which the SM is extended by a scalar electroweak multiplet with half-odd-integer weak isospin, $T = (n-1)/2$, where $n$ is an even integer that counts the number of complex fields in the multiplet. This scalar multiplet has weak hypercharge\footnote{We normalize $Y$ such that $Q = T^3 + Y/2$.} $Y=1$ (the same as that of the SM Higgs doublet), and transforms as odd under an imposed global $Z_2$ symmetry, allowing for the lightest member to be a stable dark matter candidate.  In particular, we focus on the models with weak isospin $T = 5/2$ (sextet, $n=6$) and $T = 7/2$ (octet, $n=8$).\footnote{A recent study~\cite{Hamada:2015bra} of the renormalization group running of the couplings in such models found that the sextet model's couplings will diverge at a scale below 5000~TeV, if the new scalars are at the 100~GeV scale.  The octet model was not studied explicitly in Ref.~\cite{Hamada:2015bra}, but the stronger running due to the larger electroweak representation means that its couplings will diverge at an even lower scale.  This implies either that the new scalars must be composite particles or that additional fermions must enter in such a way as to sufficiently modify the running of the scalar couplings of the large multiplet.}  Larger complex scalar multiplets are disallowed by perturbative unitarity considerations involving the scattering of two scalars into two gauge bosons~\cite{Hally:2012pu}.  Smaller multiplets, such as the $T = 3/2$ case, have previously been studied in the literature~\cite{AbdusSalam:2013eya}.

In these models, the large number of new scalar particles leads to a plethora of electroweak pair-production processes accessible at the LHC via an $s$-channel photon, $W$, or $Z$.  Because of the $Z_2$ symmetry, these scalars can decay only to lighter scalars within the large multiplet via the emission of $W$ or $Z$ bosons, terminating at the lightest state $\zo$ of the large multiplet, which is stable and escapes the detector. These production and decay modes produce similar phenomenology to supersymmetric electroweak gauginos, and the most promising searches to constrain these models are similarly those with a combination of missing transverse energy ($\met$) and leptons. We avoid signatures with too many high-energy jets in order to avoid large QCD backgrounds, such as $t \bar t$. In particular, we have recast a number of existing searches for supersymmetric electroweak gauginos by both the ATLAS \cite{TheATLAScollaboration:2013hha,ATLAS:2013rla,Aad:2014nua,ATLAS:2013qla,TheATLAScollaboration:2013via,ATLAS:2013yla,ATLAS:2013tma,ATLAS:2012zim,TheATLAScollaboration:2013aia,TheATLAScollaboration:2013fha} and CMS \cite{CMS:2013dea,1596278,CMS:rwa} collaborations which use data taken at a centre-of-mass energy of 8~TeV (LHC8).

This paper is organized as follows. In Sec.~\ref{sec:models} we introduce the two large-multiplet models first studied in Ref.~\cite{Earl:2013fpa}.  We re-express the parameter freedom in the models in terms of physical masses and mass splittings, which are most physically relevant for the kinematics of pair production and decays. We also translate the theoretical and indirect experimental constraints on the models previously studied in Ref.~\cite{Earl:2013fpa} into this new parameter space.  In Sec.~\ref{sec:decays} we discuss the pattern of decay branching fractions of each of the scalars.  In Sec.~\ref{sec:collider} we recast a collection of LHC8 searches for supersymmetric particles to constrain our models.  We conclude in Sec.~\ref{sec:conclusions}.  Some mathematical details and simulation issues are relegated to the appendices.

\section{Description of the models}
\label{sec:models}

We consider two models that extend the SM through the addition of a single, large electroweak multiplet of complex scalars, $\multiplet$, which has quantum numbers under $SU(3)_c \otimes SU(2)_L \otimes U(1)_Y$ of $(1, n, 1)$, where $n = 6$ or $8$ is the size of the multiplet. In these models, the most general gauge-invariant scalar potential that preserves a $Z_2$ symmetry under which $\multiplet \to -\multiplet$ is given by
\begin{align}
	V(\Phi,\multiplet) &= m^2 \Phi^\dag \Phi + M^2 \multiplet^\dag \multiplet + \lambda_1 \big(\Phi^\dag \Phi\big)^2 
	+ \lambda_2\Phi^\dag \Phi\, \multiplet^\dag \multiplet + \lambda_3 \Phi^\dag \tau^a\Phi\, \multiplet^\dag T^a \multiplet
	\nonumber \\
	&\hspace*{3em} + \left[\lambda_4\,\widetilde{\Phi}^\dag \tau^a \Phi\;\multiplet^\dag T^a\widetilde{\multiplet} 
	+ \mbox{h.c.}\right] + \mathcal{O}(\multiplet^4)\;,
	\label{eq:conjugatepotential}
\end{align}
where $\Phi$ is the SM $SU(2)_L$ doublet.  Here $\widetilde{\Phi} = C\Phi^*$ and $\widetilde{\multiplet} = C\multiplet^*$ are the Higgs doublet and the large scalar multiplet in the conjugate representation, respectively.
The conjugation matrix, $C$, is an antisymmetric $n \times n$ matrix equal to $i \sigma^2$ for the SU(2)$_L$ doublet, and is given in Appendix~\ref{app:massesmixing} for the $n = 6$ and $n = 8$ representations. The $\tau^a$ and $T^a$ matrices are the generators of $SU(2)_L$ in the doublet and $n$-plet representations, respectively.

The term $\multiplet^\dag T^a\widetilde{\multiplet}$ can only be non-zero when $n$ is an even number, i.e., for half-odd-integer values for the total isospin, $T$, of the scalar multiplet. Together with the constraint $T \leq 7/2$ ($n \leq 8$) for complex scalar multiplets that arises from perturbative unitarity of scattering amplitudes involving the large multiplet's weak charge~\cite{Hally:2012pu}, this limits the models of interest to the cases $T = 5/2$ ($n = 6$) and $T = 7/2$ ($n = 8$).\footnote{The model with $T = 3/2$ has been studied in Ref.~\cite{AbdusSalam:2013eya}.}  For these two cases, the large multiplet is given in the electroweak basis by
\begin{align}
	\multiplet_{(n=6)} &= \left(\zeta^{+3},\,\zeta^{+2},\,\zeta^{+1},\,\zeta^{0},\,
	\zeta^{-1},\,\zeta^{-2}\right)^T\;, \nonumber \\
	\multiplet_{(n=8)} &= \left(\zeta^{+4},\,\zeta^{+3},\,\zeta^{+2},\,\zeta^{+1},\,\zeta^{0},\,
	\zeta^{-1},\,\zeta^{-2},\,\zeta^{-3}\right)^T\;.
	\label{eq:states}
\end{align}
Note that the conjugate of the charged state $\zeta^Q$ is written as $\zeta^{Q*}$, which is not the same as $\zeta^{-Q}$.

When the $\lambda_4$ term in Eq.~(\ref{eq:conjugatepotential}) vanishes, the Lagrangian preserves a $U(1)$ symmetry (models with such a $U(1)$-symmetric potential have been studied in Ref.~\cite{Earl:2013jsa}). The $\lambda_4$ term breaks the $U(1)$ down to a $Z_2$ symmetry~\cite{Earl:2013fpa}, and splits the complex neutral component of $\multiplet$ into its real and imaginary parts, $\zo = \sqrt{2}~\mathrm{Re}~\zeta^0$ and $\zi = \sqrt{2}~\mathrm{Im}~\zeta^0$, with different masses.
Furthermore, the $\lambda_4$ term induces and controls the amount of mixing between the charged states with the same electric charge, $\zeta^{Q}$ and $\zeta^{-Q*}$. The mass eigenstates are defined for $Q > 0$ in terms of a mixing angle $\alpha_Q$ such that
\begin{align}
	H_1^{Q} &= \cos \alpha_Q \, \zeta^{Q} + \sin \alpha_Q \, \zeta^{-Q*}\;,\nonumber\\
	H_2^{Q} &= -\sin \alpha_Q \, \zeta^{Q} + \cos \alpha_Q \, \zeta^{-Q*}\;,
	\label{eq:alphadef}
\end{align}
with $m_{H_1^Q} < m_{H_2^Q}$. Since there is only one state with $|Q| = n/2$, it remains unmixed. Expressions for the mixing angles are given in Eqs. (7) and (14) of Ref.~\cite{Earl:2013fpa}, and the details are summarized in Appendix~\ref{app:massesmixing} for completeness.\footnote{We correct a typographical error in Eq.~(A7) of Ref.~\cite{Earl:2013fpa} for the generic expression for the mixing angle $\alpha_Q$.  The expressions in Eqs.~(7) and (14) of Ref.~\cite{Earl:2013fpa} are correct.}

For these models to contain a dark matter candidate, we require that the lightest (stable) member of the large multiplet be electrically neutral. This occurs only when $|\lambda_3| < 2|\lambda_4|$.  In addition, and without loss of generality, we choose the real part of $\zeta^0$ to be the lightest member of the large multiplet; this constrains the sign of $\lambda_4$ such that $\lambda_4 < 0$ for the sextet model and $\lambda_4 > 0$ for the octet.

\begin{figure}
\centering
	\includegraphics[width=0.24\textwidth]{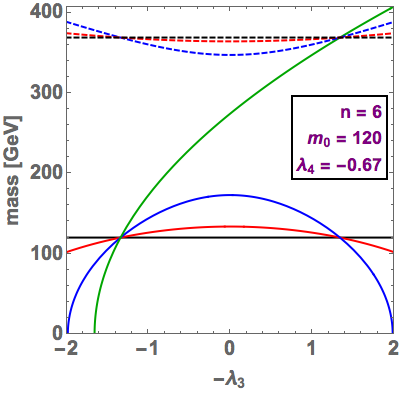}
	\includegraphics[width=0.24\textwidth]{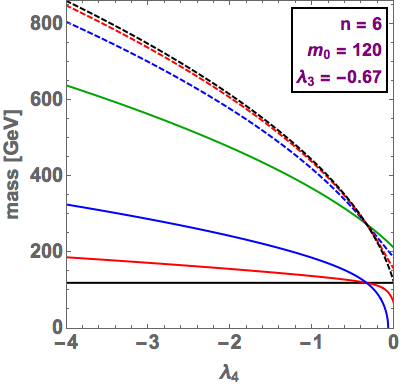}
	\includegraphics[width=0.24\textwidth]{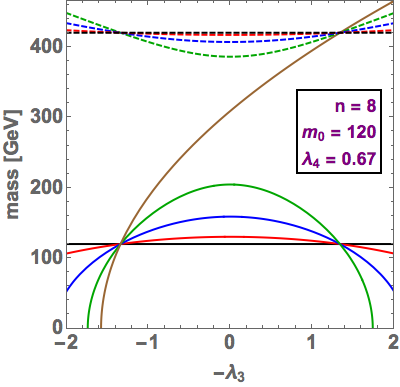}
	\includegraphics[width=0.24\textwidth]{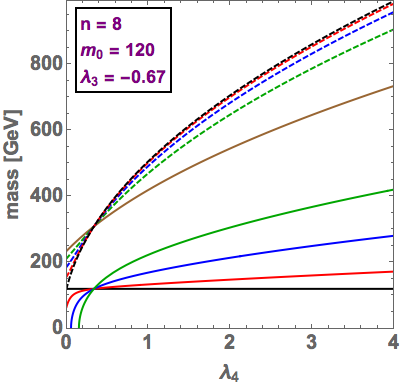}
	\caption{Sample mass spectra for fixed values of $\lambda_3$ or $\lambda_4$. The left two plots are for the $n=6$ model and show, from top to bottom, $\zeta^{0,i}$ (dashed black), $H_2^{+}$ (dashed red), $H_2^{++}$ (dashed blue), $\zeta^{+3}$ (solid green), $H_1^{++}$ (solid blue), $H_1^{+}$ (solid red), and $\zeta^{0,r}$ (solid black). The right two plots are for the $n=8$ model and show, from top to bottom, $\zeta^{0,i}$ (dashed black), $H_2^{+}$ (dashed red), $H_2^{++}$ (dashed blue), $H_2^{+3}$ (dashed green), $\zeta^{+4}$ (solid brown), $H_1^{+3}$ (solid green), $H_1^{++}$ (solid blue), $H_1^{+}$ (solid red), and $\zeta^{0,r}$ (solid black).}
	\label{fig:massspectra}
\end{figure}

The masses of the physical states in terms of the mass of the neutral real particle, $\mr$, and the Lagrangian parameters $\lambda_3$ and $\lambda_4$ are given by~\cite{Earl:2013fpa}
\begin{eqnarray}
	\mi^2 &=& \mr^2 + \frac{n}{2}(-1)^{\frac{n}{2}}v^2\lambda_4, \nonumber \\
	m_{H_{1,2}^{+Q}}^2 &=& \mr^2 + \frac{1}{4}v^2\left[n(-1)^{\frac{n}{2}}\lambda_4 \mp \sqrt{Q^2\lambda_3^2 + (n^2-4Q^2)\lambda_4^2}\right], \nonumber \\
	m_{\zeta^{+\frac{n}{2}}}^2 &=& \mr^2 - \frac{n}{8}v^2\left[\lambda_3 + 2 (-1)^{\frac{n}{2}+1}\lambda_4\right],
	\label{eq:masses} 
\end{eqnarray}
where the notation is such that the sign in $m_{H_{1,2}^Q}^2$ forces the relation $m_{H_1^Q} < m_{H_2^Q}$.  The physical scalars arising from the large multiplet always occur in the same hierarchy, given from lightest to heaviest by:
\begin{eqnarray}
	\multiplet_{(n=6)} &\to & \zeta^{0,r}, H_1^{\pm}, H_1^{\pm\pm}, \zeta^{\pm 3}, H_2^{\pm\pm}, H_2^{\pm}, \zeta^{0,i}, \nonumber \\
	\multiplet_{(n=8)} &\to &  \zeta^{0,r}, H_1^{\pm}, H_1^{\pm\pm}, H_1^{\pm 3}, \zeta^{\pm 4}, H_2^{\pm 3}, H_2^{\pm\pm}, H_2^{\pm}, \zeta^{0,i}.
\end{eqnarray}
Sample mass spectra are shown in Fig.~\ref{fig:massspectra}, where the two left plots are for $n = 6$ and the two right plots are for $n = 8$. For each of the two plots for each multiplet, we have fixed one of $\lambda_3$ or $\lambda_4$ and varied the other to illustrate the effect of varying these parameters.

\begin{figure}
\centering
	\includegraphics[width=0.4\textwidth]{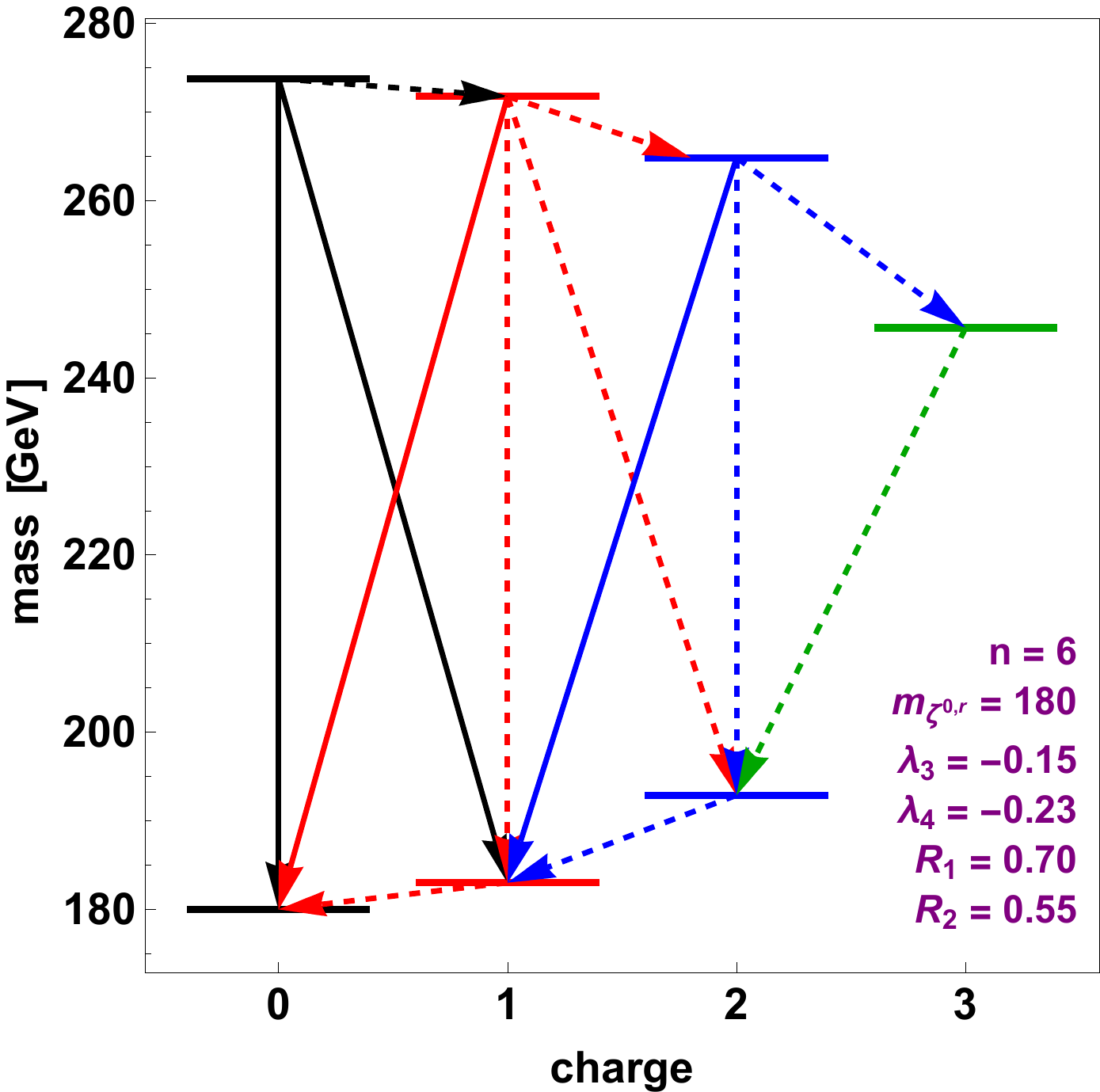}
	\hspace*{1em}
	\includegraphics[width=0.4\textwidth]{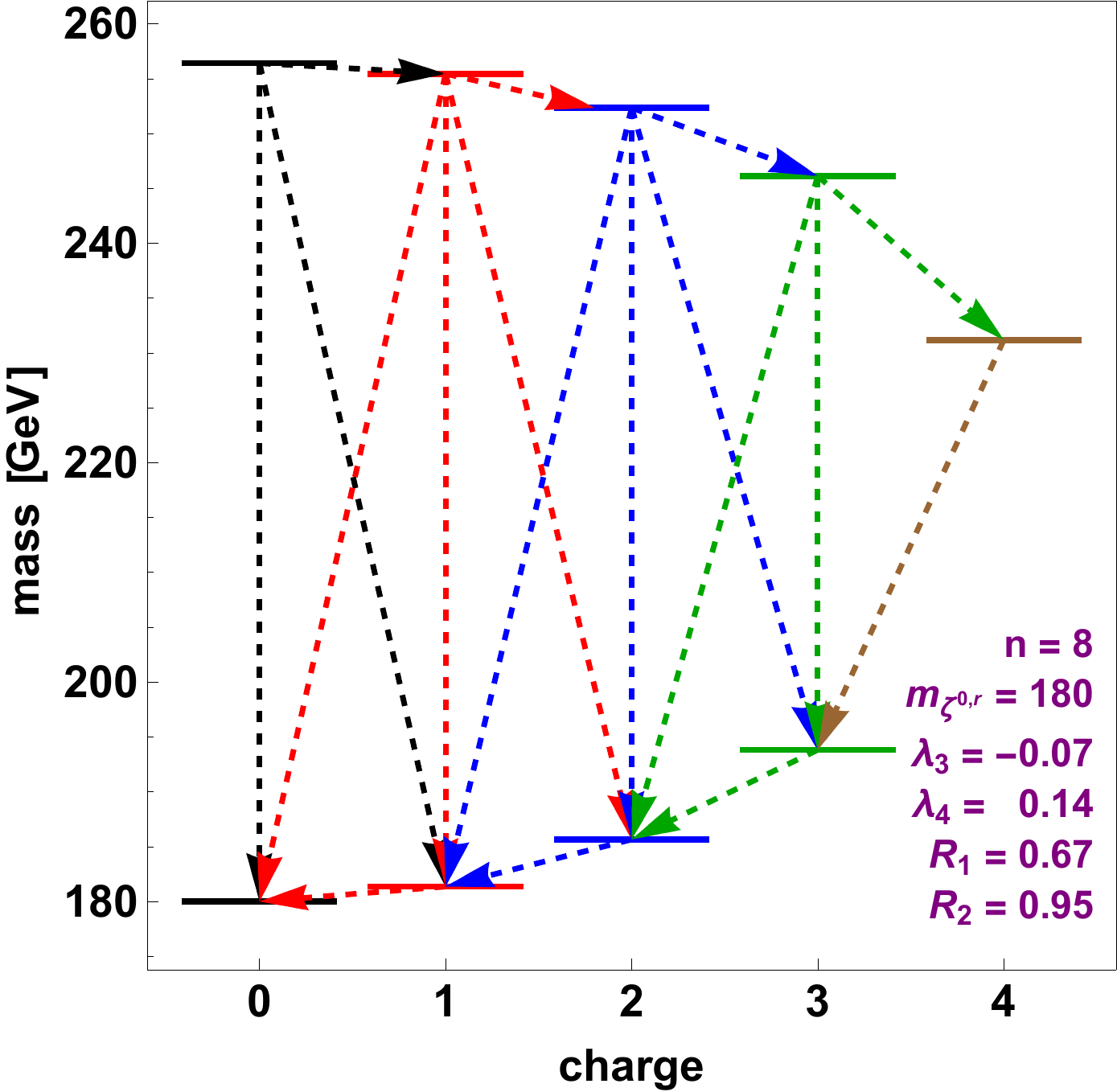}
	\caption{The decay patterns of the $n=6$ (left) and $n=8$ (right) models for a typical parameter point. The upper branch of states includes the $\zi$, $H_2^{+Q}$ and $\zeta^{+\frac{n}{2}}$, while the lower branch includes $\zo$ and $H_1^{+Q}$, each sorted by electric charge. Decays are shown as diagonal ($W$ emission) or vertical ($Z$ emission) arrows. On-shell decays are indicated by a solid line and off-shell decays by a dashed line.}
	\label{fig:chargespectrum}
\end{figure}

As seen in Eq.~(\ref{eq:conjugatepotential}), the states of $\multiplet$ interact with the SM through a coupling to the Higgs doublet, as well as through their gauge-kinetic terms. The couplings to the photon and the Higgs boson are diagonal in the mass basis, and thus do not induce decays. However, off-diagonal vertices involving two different scalar mass eigenstates and a $Z$ or $W$ boson do exist. Thus, the decays of the mass eigenstates in $\multiplet$ to lighter members of $\multiplet$ occur only through emission of a $W$ or $Z$, which may be off-shell depending on the mass splitting involved.  This leads to a distinctive decay pattern, which is shown for the $n=6$ and $n=8$ models in Fig.~\ref{fig:chargespectrum} for a typical parameter point.

Reference~\cite{Earl:2013fpa} studied theoretical and indirect experimental constraints on the model parameters $(\mr,\,\lambda_2,\,\lambda_3,\,\lambda_4)$, which we apply here:
\begin{itemize}
\item[$(i.)$] The zeroth partial wave $2 \to 2$ scattering amplitude involving pairs of scalars must satisfy perturbative unitarity, $|\mathrm{Re}\,a_0| \leq 1/2$. This results in constraints for the sextet model of $|\lambda_2| \leq$ 6.59, $|\lambda_3| \leq$ 8.48, and $|\lambda_4| \leq$ 4.25, and for the  
octet model of $|\lambda_2| \leq$ 3.10, $|\lambda_3| \leq$ 5.46, and $|\lambda_4| \leq$ 2.74. 
\item[$(ii.)$] The scalar potential should not have an alternative minimum in which $\multiplet$ gets a vacuum expectation value.  A sufficient condition to ensure this is to set $M^2 > 0$ in Eq.~(\ref{eq:conjugatepotential}). 
\item[$(iii.)$] The electroweak oblique parameters $S$, $T$, and $U$~\cite{Peskin:1990zt}, receive contributions from electroweak gauge boson self-energy diagrams involving the states in $\multiplet$. The measured oblique parameters constrain $|\lambda_3| \approx 2|\lambda_4|$ and favour $\lambda_3 \lesssim 0$. This constraint causes the $\zeta^{\frac{n}{2}}$ state to tend to cluster with the heavier scalars, as can be seen in the first and third plots in Fig.~\ref{fig:massspectra}.
This will be the case unless the overall spectrum is highly compressed. 
\item[$(iv.)$] The SM Higgs boson decay width to two photons receives contributions from the charged scalars in $\multiplet$ running in the loop.  The LHC measurement of the $h \to \gamma\gamma$ rate constrains $\lambda_2$ more strongly than the unitarity constraint when $\mr \lesssim 500$ GeV. Since $\lambda_2$ does not appear explicitly in the mass formulae in Eq.~(\ref{eq:masses}) or affect any of the electroweak production and decay rates, we satisfy this constraint by setting $\lambda_2 = 0$ for this work.  We checked numerically that this choice does not further restrict the range of the remaining parameters beyond the previous three constraints.
\end{itemize}

In this study, it is most convenient to reparameterize the models in terms of physical masses and mass splittings.  We take as inputs $\mr$ as in Eq.~(\ref{eq:masses}), and replace $\lambda_3$ and $\lambda_4$ with the mass splitting parameters $R_1$ and $R_2$, defined as
\begin{align}
	R_1 &\equiv \frac{\Delta m_{\frac{n}{2}}}{\Delta m_{0}} = \frac{m_{\zeta^{+\frac{n}{2}}} - \mr}{\mi - \mr}\;,&
	R_2 &\equiv \frac{\Delta m_{0}}{\Delta m_{0}^{\mathrm{MAX}}} = \frac{\mi - \mr}{(\mi - \mr)^{\mathrm{MAX}}}.
	\label{eq:R1R2def}
\end{align}
Here $R_1$ parameterizes the mass of the highest-charged state, $\zeta^{n/2}$, in terms of the mass  splitting $\Delta m_{\frac{n}{2}} \equiv m_{\zeta^{+\frac{n}{2}}} - \mr$ as a fraction of the mass splitting between the lightest and heaviest state of $\multiplet$, $\Delta m_0 \equiv \mi - \mr$ (see Fig.~\ref{fig:massspectra}), while
$R_2$ parameterizes $\Delta m_0$ as a fraction of the maximum such splitting $\DmMAX$ allowed after the theoretical and indirect experimental constraints are imposed on the model.
The numerical values of $\DmMAX$ allowed by the theoretical and indirect experimental constraints are given in Table~\ref{tbl:Dm0MAX} for the four values of $\mr$ used in our simulation.
We study four mass slices, with $\mr = 80$, 120, 180, and 300~GeV.  (LHC8 does not provide any exclusions for $\mr$ larger than 300~GeV in these models.)

\begin{table}
\begin{center}
\begin{tabular}{C{6em}C{4.5em}C{4.5em}}
\hline\hline
\multicolumn{1}{c}{}&\multicolumn{2}{c}{$\Delta m_{0}^{\mathrm{MAX}}$ [GeV]}\\
\multicolumn{1}{c}{$\mr$ [GeV]} & $(n = 6)$ & $(n = 8)$\\
\hline
$80$ & $76.98$ & $39.29$\\
$120$ & $114.38$ & $55.75$\\
$180$ & $170.48$ & $80.43$\\
$300$ & $282.68$ & $129.81$\\
\hline \hline
\end{tabular}
\caption{Maximum mass splitting between $\zeta^{0,r}$ and $\zeta^{0,i}$, $\DmMAX \equiv \mi - \mr$, for particular values of $\mr$ in the two models. Values were obtained numerically by applying the theoretical and indirect experimental constraints described in Sec.~\ref{sec:models}.}
\label{tbl:Dm0MAX}
\end{center}
\end{table}

The mass splitting parameters are normalized to fall in the ranges $R_1 \in [0,\,1]$ and $R_2 \in [0,\,1]$. From Eq.~(\ref{eq:masses}), it is clear that as $\lambda_3 \to +2|\lambda_4|$, $R_1 \to 0$, and as $\lambda_3 \to -2|\lambda_4|$, $R_1 \to 1$. Furthermore, as illustrated in Fig.~\ref{fig:massspectra}, the mass splittings among the states are maximized when $R_1 \sim 0.5$, while the spectrum collapses into two tightly-clustered groups of states when $R_1 \to 0$ or 1. Also, because $R_2$ parameterizes the overall mass splitting between the neutral real and imaginary scalars, the entire spectrum becomes compressed as $R_2 \to 0$, whereas the splitting between the heavier and lighter states is maximized when $R_2 \to 1$.  

\begin{figure}
\centering
	\includegraphics[width=0.24\textwidth]{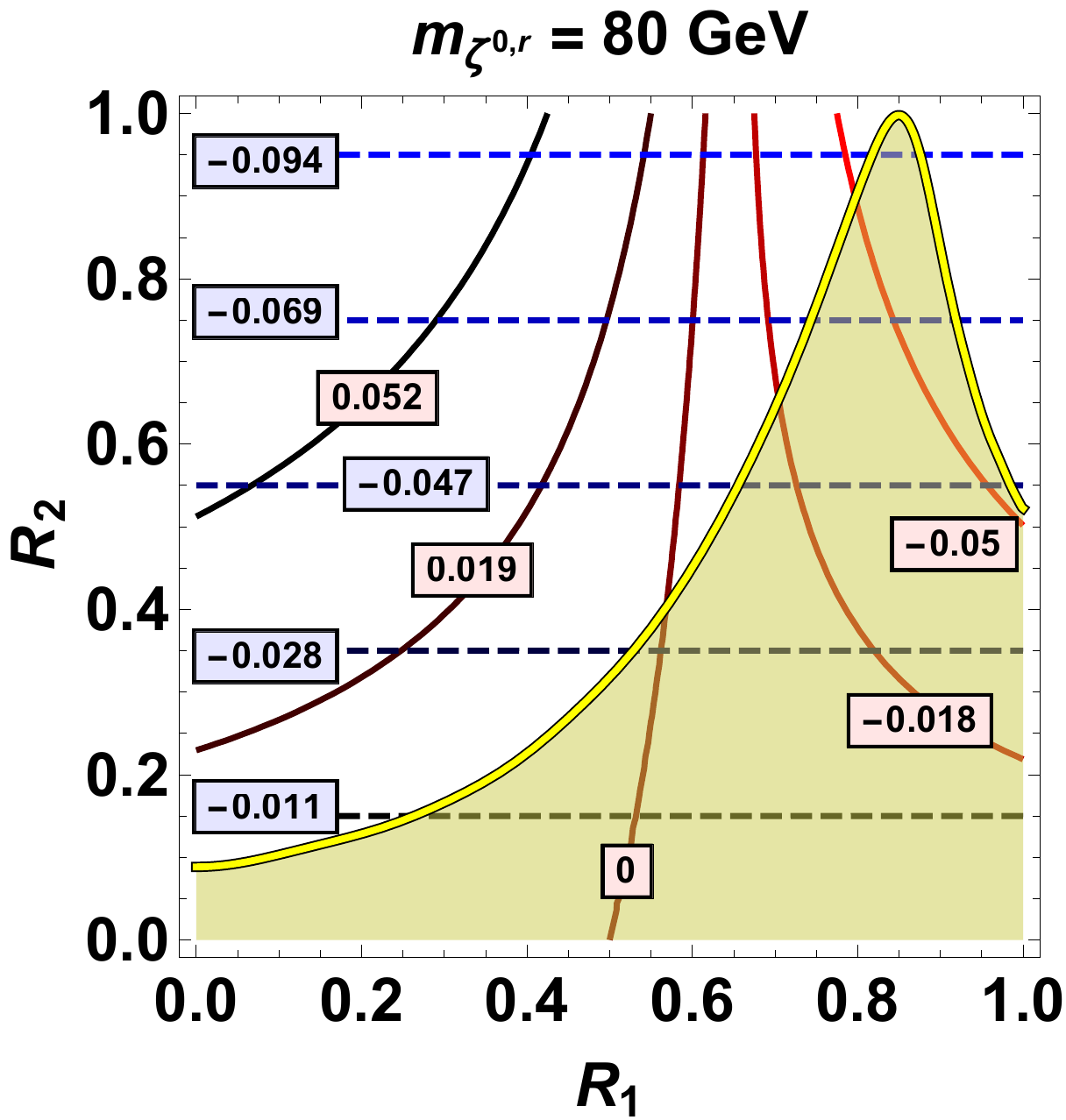}
	\includegraphics[width=0.24\textwidth]{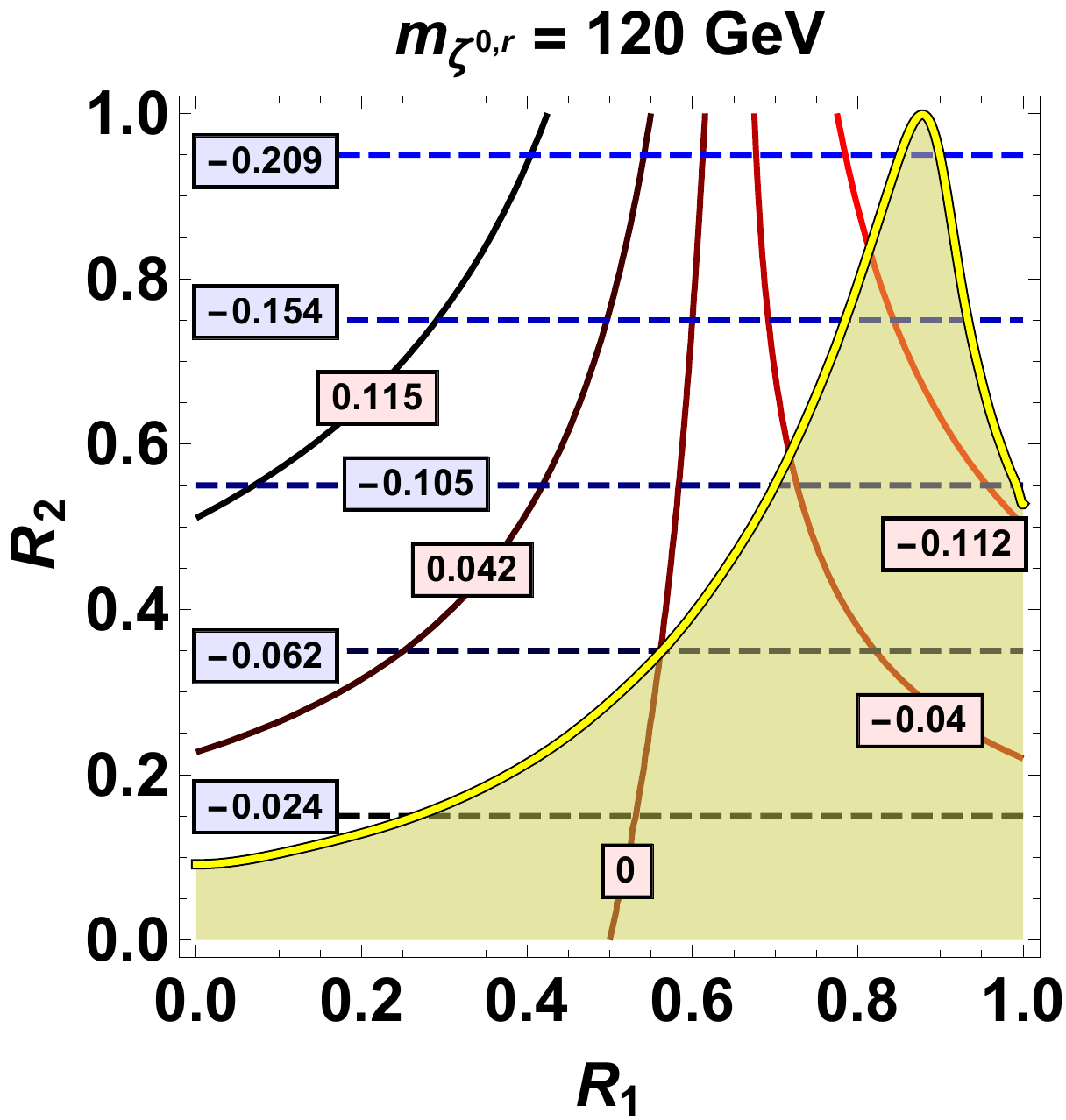}
	\includegraphics[width=0.24\textwidth]{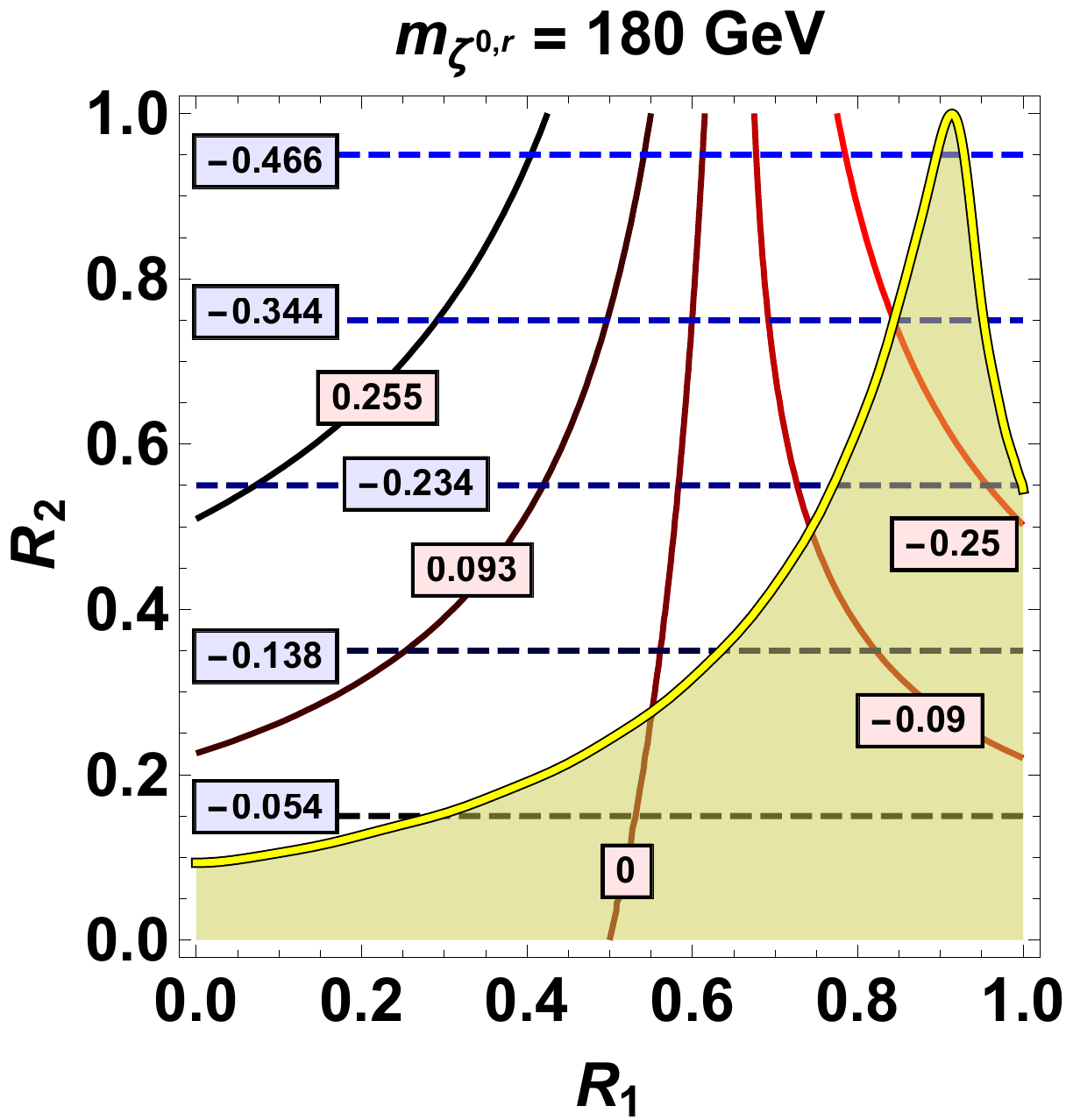}
	\includegraphics[width=0.24\textwidth]{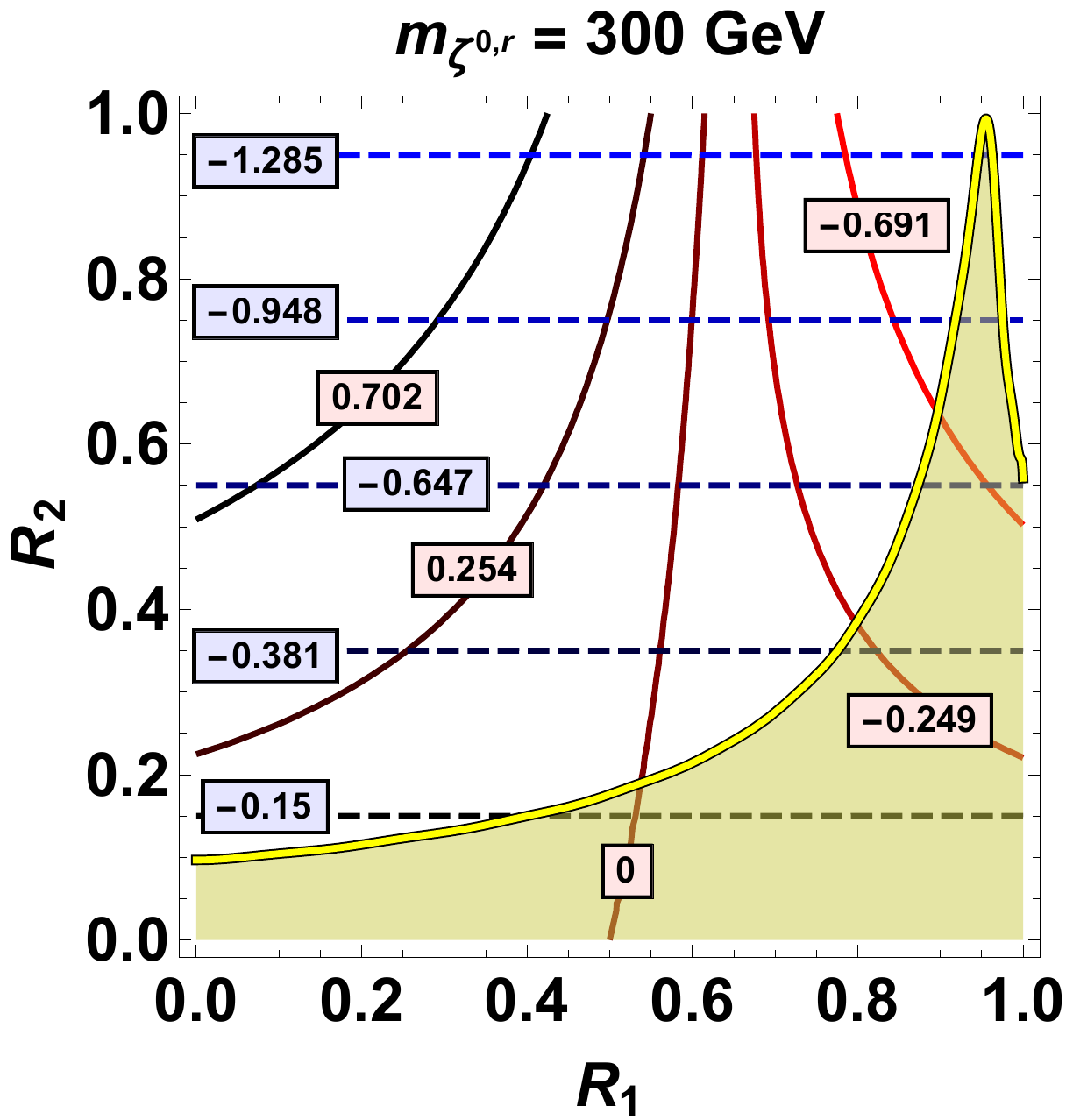}
	\caption{Parameter region allowed by the theoretical and indirect experimental constraints from Ref.~\cite{Earl:2013fpa} in the $R_1$--$R_2$ plane, for four values of $\mr$ in the $n=6$ model. The allowed region is shaded in yellow. Contours of constant $\lambda_3$ are shown as solid curves with pink labels and contours of constant $\lambda_4$ are shown as dashed horizontal lines with blue labels. Curves are labelled with their respective $\lambda_3$ or $\lambda_4$ value.}
	\label{fig:R1R2envconstL3L4_n6}
\end{figure}

\begin{figure}
\centering
	\includegraphics[width=0.24\textwidth]{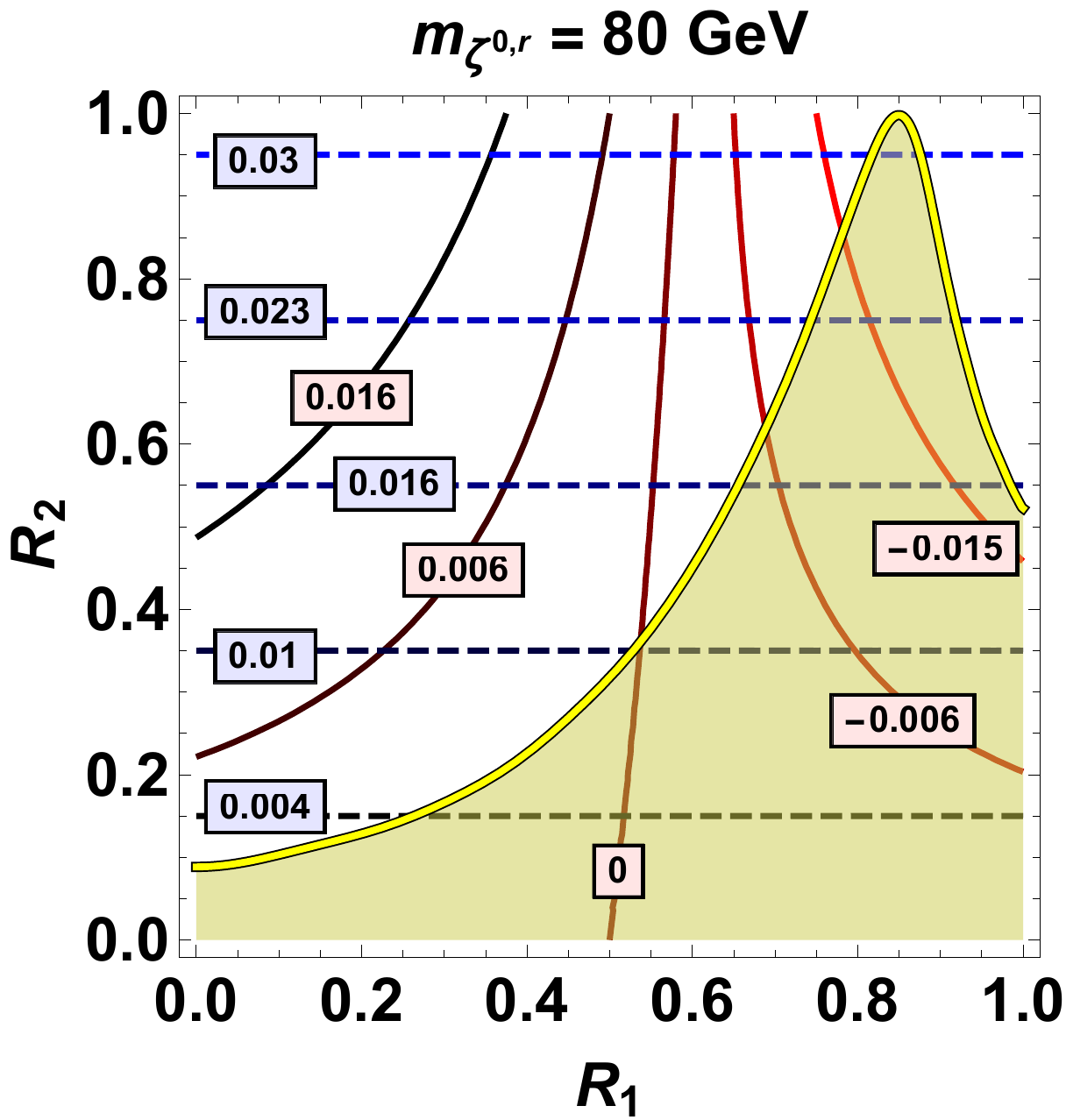}
	\includegraphics[width=0.24\textwidth]{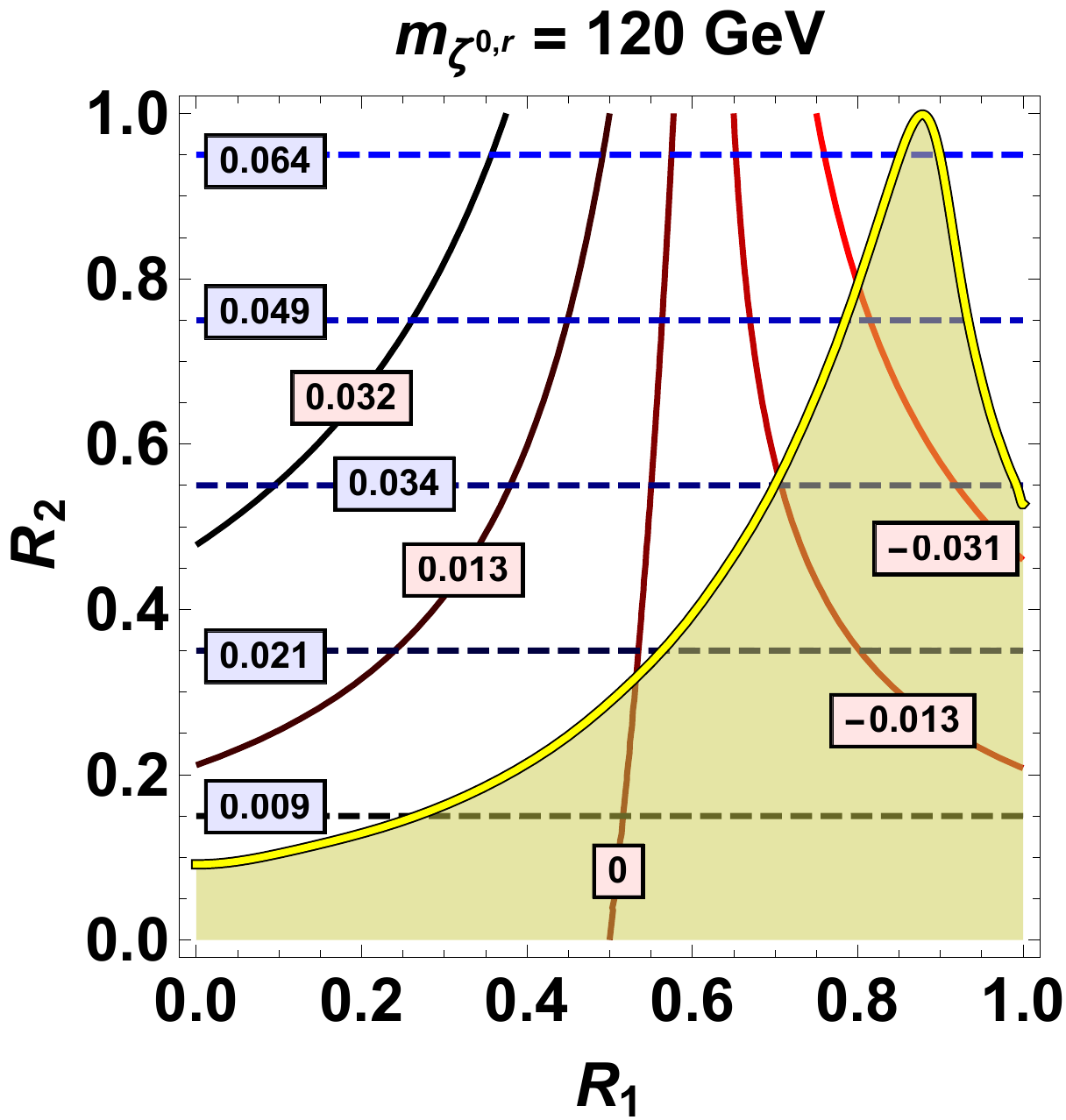}
	\includegraphics[width=0.24\textwidth]{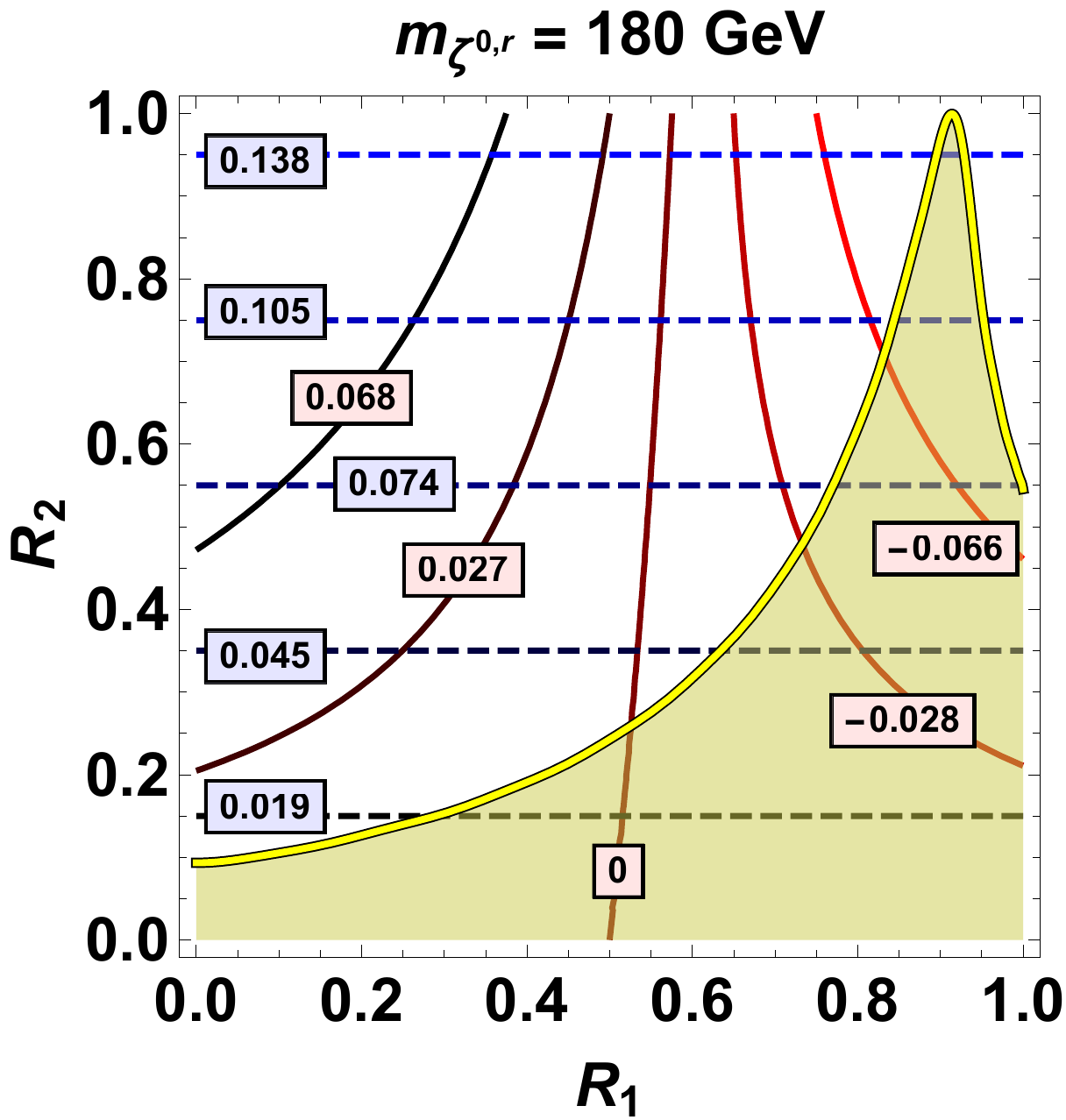}
	\includegraphics[width=0.24\textwidth]{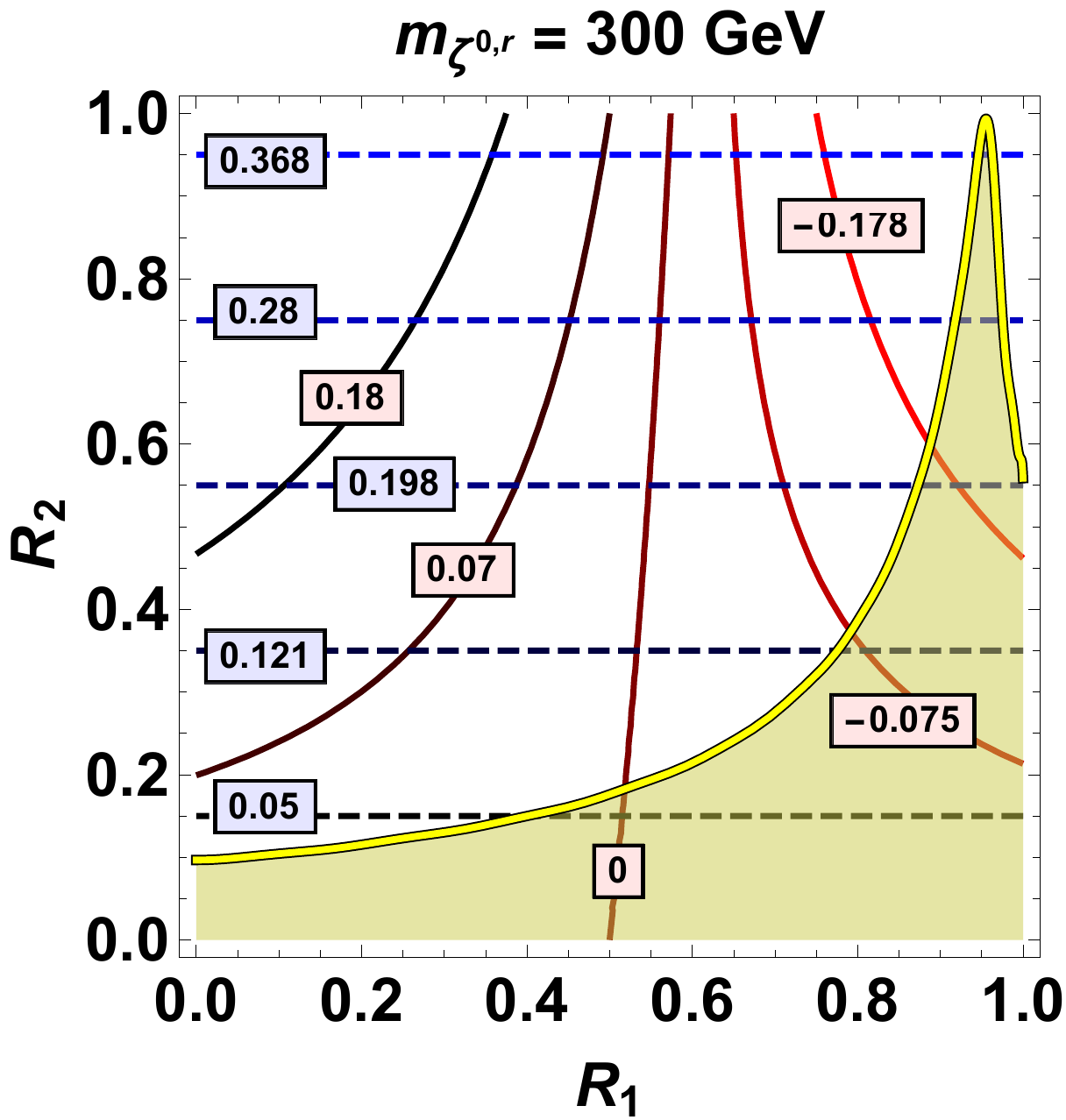}
	\caption{As in Fig.~\ref{fig:R1R2envconstL3L4_n6} but for the $n=8$ model.}
	\label{fig:R1R2envconstL3L4_n8}
\end{figure}

The parameter regions allowed by the theoretical and indirect experimental constraints studied in Ref.~\cite{Earl:2013fpa} are plotted in Figs.~\ref{fig:R1R2envconstL3L4_n6} and \ref{fig:R1R2envconstL3L4_n8} in the $R_1$--$R_2$ plane, for the four $\mr$ values that we study.  The yellow shaded regions are allowed.  On each plot we also show the contours of constant $\lambda_3$ and $\lambda_4$.

\section{Preferred decay modes}
\label{sec:decays}

\begin{figure}
\centering
	\includegraphics[width=0.24\textwidth]{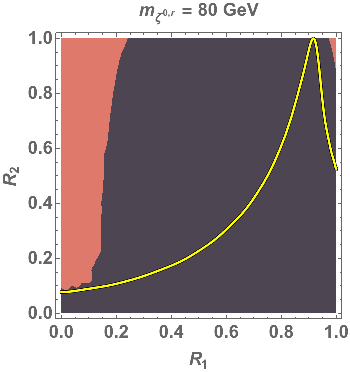}
	\includegraphics[width=0.24\textwidth]{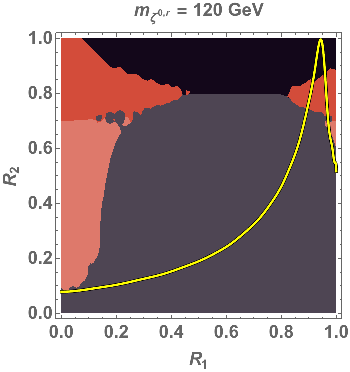}
	\includegraphics[width=0.24\textwidth]{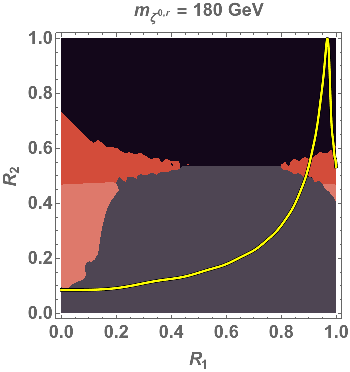}
	\includegraphics[width=0.24\textwidth]{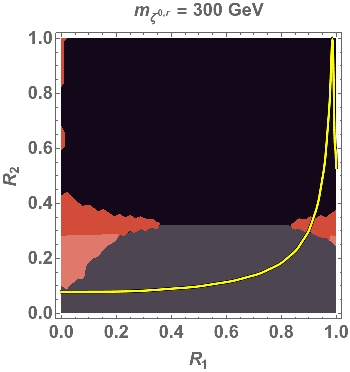}
	
	\includegraphics[height=2.24em]{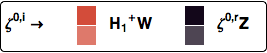}
	\caption{Dominant decay modes of $\zeta^{0,i}$ in the $n = 6$ model.  Shown are $\zeta^{0,i} \to H_1^{\pm} W^{\mp}$ (orange/red) and $\zeta^{0,i} \to \zeta^{0,r} Z$ (purple/black).  The decay $\zeta^{0,i} \to H_2^{\pm} W^{\mp}$ is severely kinematically suppressed over the entire parameter space and never dominates.  The darker shade of each colour indicates that the gauge boson is emitted on-shell, while the lighter shade indicates it is off-shell.  The region below the yellow curve is allowed by indirect and precision electroweak constraints.}
	\label{fig:decays_Z0I_n6}
\end{figure}

\begin{figure}
\centering
	\includegraphics[width=0.24\textwidth]{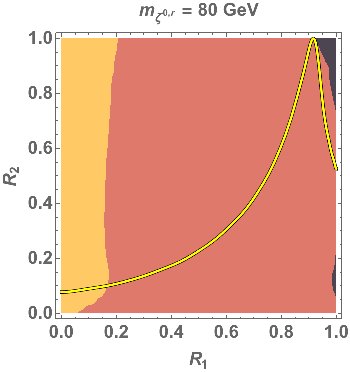}
	\includegraphics[width=0.24\textwidth]{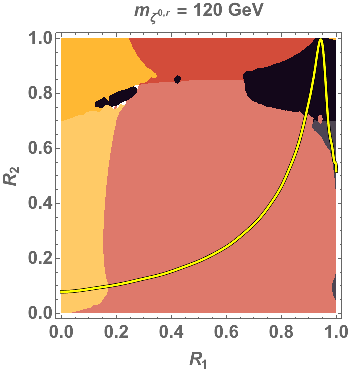}
	\includegraphics[width=0.24\textwidth]{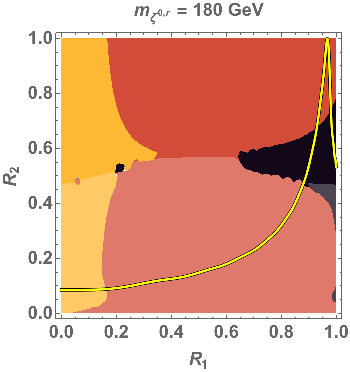}
	\includegraphics[width=0.24\textwidth]{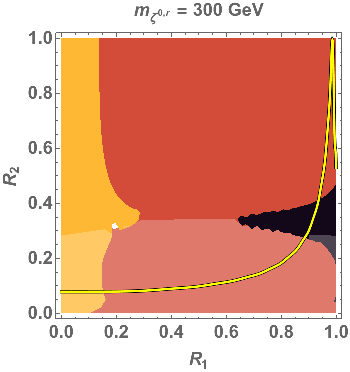}
	
	\includegraphics[height=2.24em]{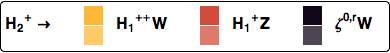}
	\caption{Dominant decay modes of $H_2^{+}$ in the $n = 6$ model. Shown are $H_2^+ \to H_1^{++} W^-$ (yellow), $H_2^+ \to H_1^+ Z$ (orange/red), and $H_2^+ \to \zeta^{0,r} W^+$ (purple/black).  The decay $H_2^+ \to H_2^{++} W^-$ is severely kinematically suppressed over the entire parameter space and never dominates.  The darker shade of each colour indicates that the gauge boson is emitted on-shell, while the lighter shade indicates it is off-shell. The region below the yellow curve is allowed by indirect and precision electroweak constraints.}
	\label{fig:decays_H21_n6}
\end{figure}

\begin{figure}
\centering
	\includegraphics[width=0.24\textwidth]{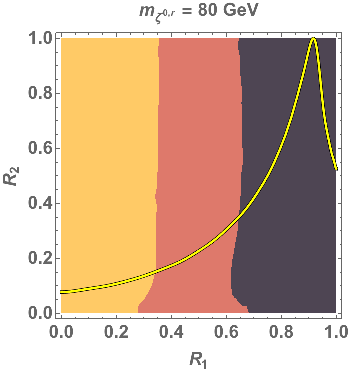}
	\includegraphics[width=0.24\textwidth]{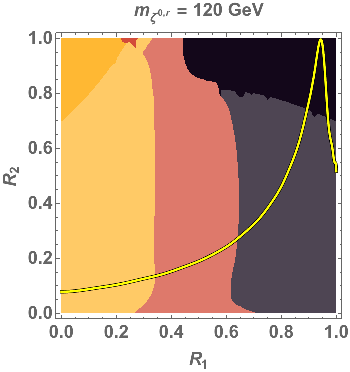}
	\includegraphics[width=0.24\textwidth]{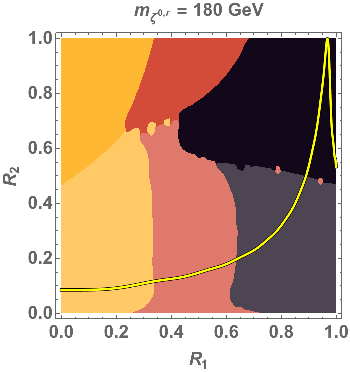}
	\includegraphics[width=0.24\textwidth]{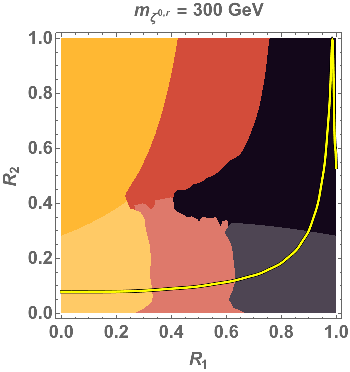}
	
	\includegraphics[height=2.24em]{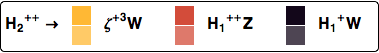}
	\caption{Dominant decay modes of $H_2^{++}$ in the $n = 6$ model.  Shown are $H_2^{++} \to \zeta^{+3} W^-$ (yellow), $H_2^{++} \to H_1^{++} Z$ (orange/red), and $H_2^{++} \to H_1^+ W^+$ (purple/black). The darker shade of each colour indicates that the gauge boson is emitted on-shell, while the lighter shade indicates it is off-shell. The region below the yellow curve is allowed by indirect and precision electroweak constraints.}
	\label{fig:decays_H22_n6}
\end{figure}

\begin{figure}
\centering
	\includegraphics[width=0.24\textwidth]{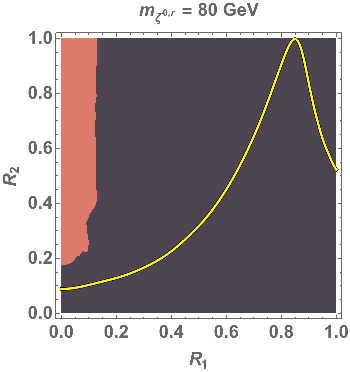}
	\includegraphics[width=0.24\textwidth]{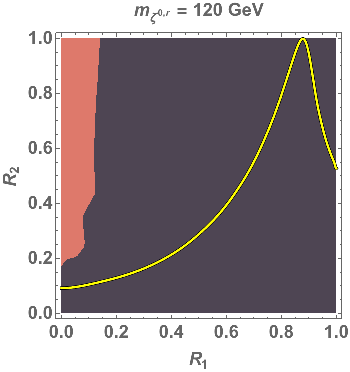}
	\includegraphics[width=0.24\textwidth]{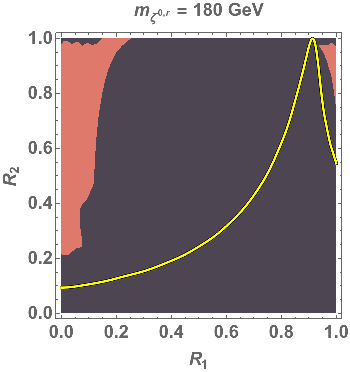}
	\includegraphics[width=0.24\textwidth]{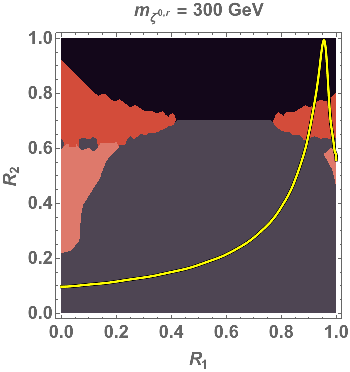}
	
	\includegraphics[height=2.24em]{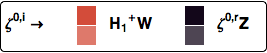}
	\caption{Dominant decay modes of $\zeta^{0,i}$ in the $n = 8$ model.  Shown are $\zeta^{0,i} \to H_1^{\pm} W^{\mp}$ (orange/red) and $\zeta^{0,i} \to \zeta^{0,r} Z$ (purple/black). The decay $\zeta^{0,i} \to H_2^{\pm} W^{\mp}$ is severely kinematically suppressed over the entire parameter space and never dominates.  The darker shade of each colour indicates that the gauge boson is emitted on-shell, while the lighter shade indicates it is off-shell. The region below the yellow curve is allowed by indirect and precision electroweak constraints.}
	\label{fig:decays_Z0I_n8}
\end{figure}

\begin{figure}
\centering
	\includegraphics[width=0.24\textwidth]{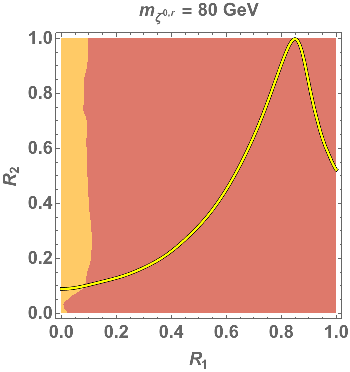}
	\includegraphics[width=0.24\textwidth]{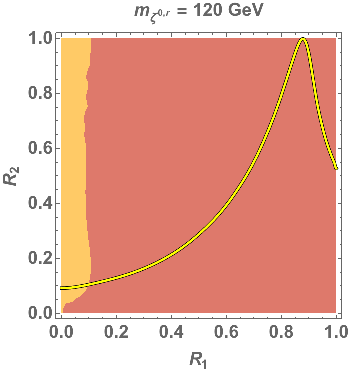}
	\includegraphics[width=0.24\textwidth]{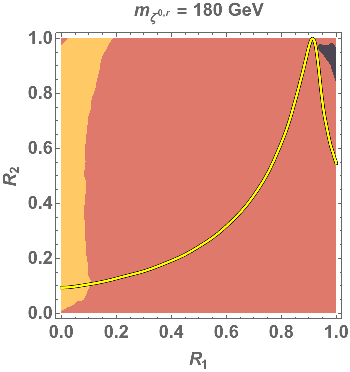}
	\includegraphics[width=0.24\textwidth]{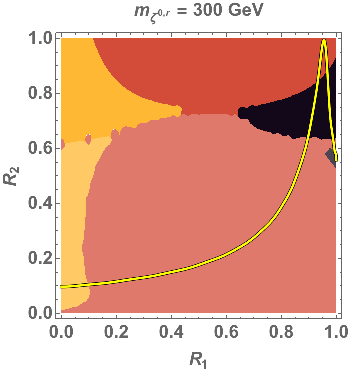}
	
	\includegraphics[height=2.24em]{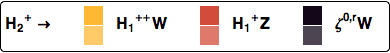}
	\caption{Dominant decay modes of $H_2^{+}$ in the $n = 8$ model.  Shown are $H_2^+ \to H_1^{++} W^-$ (yellow), $H_2^+ \to H_1^+ Z$ (orange/red), and $H_2^+ \to \zeta^{0,r} W^+$ (purple/black).  The decay $H_2^+ \to H_2^{++} W^-$ is severely kinematically suppressed over the entire parameter space and never dominates. The darker shade of each colour indicates that the gauge boson is emitted on-shell, while the lighter shade indicates it is off-shell. The region below the yellow curve is allowed by indirect and precision electroweak constraints.}
	\label{fig:decays_H21_n8}
\end{figure}

\begin{figure}
\centering
	\includegraphics[width=0.24\textwidth]{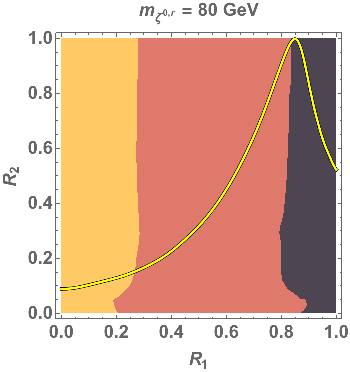}
	\includegraphics[width=0.24\textwidth]{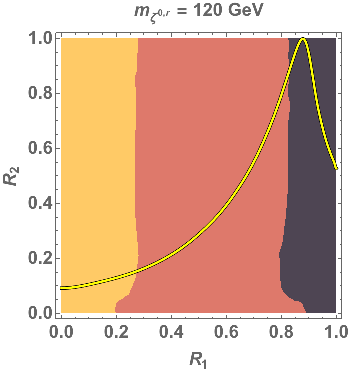}
	\includegraphics[width=0.24\textwidth]{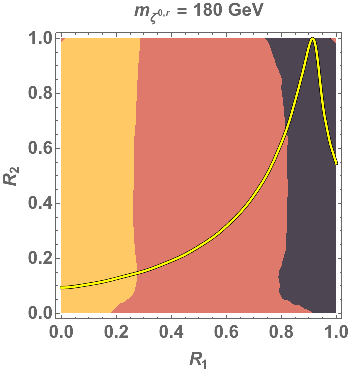}
	\includegraphics[width=0.24\textwidth]{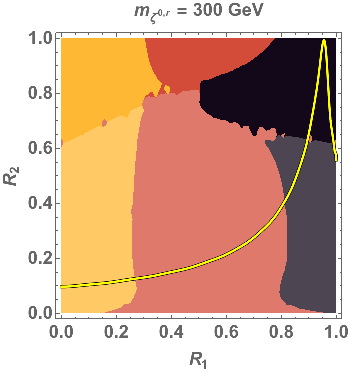}
	
	\includegraphics[height=2.24em]{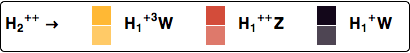}
	\caption{Dominant decay modes of $H_2^{++}$ in the $n = 8$ model.  Shown are $H_2^{++} \to H_1^{+3} W^-$ (yellow), $H_2^{++} \to H_1^{++} Z$ (orange/red), and $H_2^{++} \to H_1^+ W^+$ (purple/black).  The decay $H_2^{++} \to H_2^{+3} W^-$ is severely kinematically suppressed over the entire parameter space and never dominates. The darker shade of each colour indicates that the gauge boson is emitted on-shell, while the lighter shade indicates it is off-shell. The region below the yellow curve is allowed by indirect and precision electroweak constraints.}
	\label{fig:decays_H22_n8}
\end{figure}

\begin{figure}
\centering
	\includegraphics[width=0.24\textwidth]{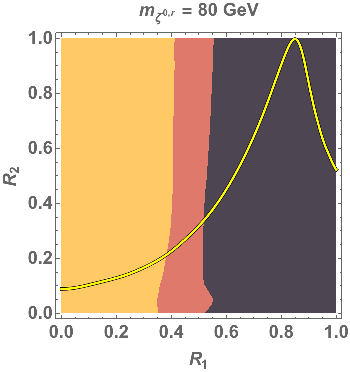}
	\includegraphics[width=0.24\textwidth]{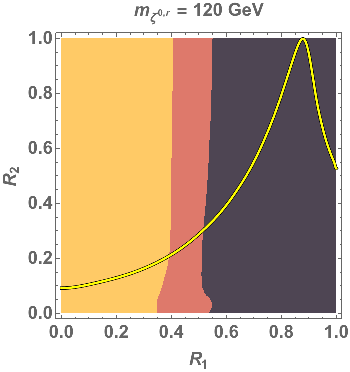}
	\includegraphics[width=0.24\textwidth]{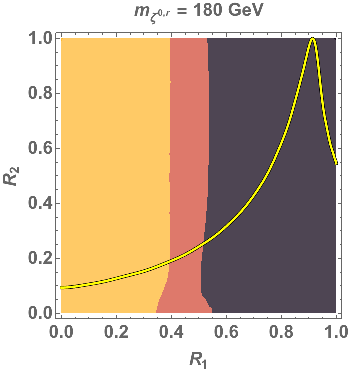}
	\includegraphics[width=0.24\textwidth]{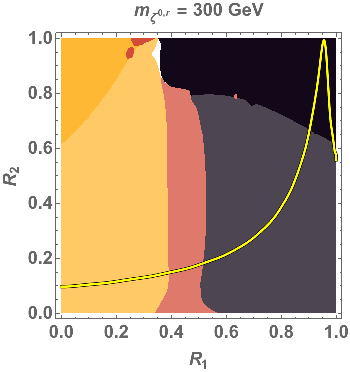}
	
	\includegraphics[height=2.24em]{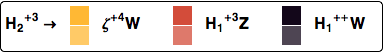}
	\caption{Dominant decay modes of $H_2^{+3}$ in the $n = 8$ model.  Shown are $H_2^{+3} \to \zeta^{+4} W^-$ (yellow), $H_2^{+3} \to H_1^{+3} Z$ (orange/red), and $H_2^{+3} \to H_1^{++} W^-$ (purple/black). The darker shade of each colour indicates that the gauge boson is emitted on-shell, while the lighter shade indicates it is off-shell. The region below the yellow curve is allowed by indirect and precision electroweak constraints.}
	\label{fig:decays_H23_n8}
\end{figure}

The mass splittings and preferred decay modes between the physical scalars in our models vary substantially over the allowed parameter space. Understanding the collider constraints requires an understanding of the branching fractions of the unstable states over the $R_1$--$R_2$ plane. In this section, we provide insight into the experimental results through examining the primary decay modes of the heavier branch of states over the full $R_1$--$R_2$ plane, for $\mr = 80$, 120, 180, and 300~GeV.  Decays in the $n=6$ model are shown in Figs.~\ref{fig:decays_Z0I_n6}--\ref{fig:decays_H22_n6} and decays in the $n=8$ model are shown in Figs.~\ref{fig:decays_Z0I_n8}--\ref{fig:decays_H23_n8}.  We do not plot the decay branching fractions of the lighter states, as these states always cascade down to $\zeta^{0,r}$ through $W$ emission as illustrated in Fig.~\ref{fig:chargespectrum}. 

Except for small regions in the transition from one dominant decay mode to another, the dominant mode has a branching fraction of at least 50\%.  We note that decays of a heavy state to another state in the heavier branch, such as $\zeta^{0,i} \to H_2^+ W$, are never dominant.  This is because they are highly suppressed by the lack of available phase space, making them effectively negligible over the entire parameter space.

Our numerical results were obtained as follows.
We implemented the two models in \texttt{FeynRules} version~2.0.26~\cite{Alloul:2013bka} and generated the corresponding Universal FeynRules Output (UFO) files.  We used these with \texttt{MadGraph5\_aMC@NLO} version~2.1.2~\cite{Alwall:2014hca,Alwall:2014bza} to compute the decay branching ratios of all the new scalars in a grid over the $R_1$--$R_2$ plane for each of the four $\mr$ slices.  

The branching ratios vary over the parameter space due to variation of the scalar mass splittings and the mixing angles $\alpha_Q$, which in turn control the gauge couplings involved in the decays.  The variation in the dominant decay mode with $R_1$ is primarily due to variation in the decay couplings, which leads to the essentially vertical stripes in Figs.~\ref{fig:decays_Z0I_n6}--\ref{fig:decays_H23_n8}.  The dependence on the mass splittings appears in regions of parameter space where one or more decays are transitioning between off-shell and on-shell.  For example, in Fig.~\ref{fig:decays_Z0I_n6}, the $\zeta^{0,i} \to H_1^{\pm} W^{\mp}$ decay goes on-shell at a lower $R_2$ value than the other decay $\zeta^{0,i} \to \zeta^{0,r} Z$, leading the former to encroach into the parameter region that would have otherwise been dominated by the latter.
This effect also shows up in Fig.~\ref{fig:decays_H21_n6}: because $\mr < m_{H_1^{++}}$, the decay of $H_2^+$ to $\zeta^{0,r} W^+$ goes on-shell while the decay to $H_1^{++} W^-$ is still off-shell, leading to a small region of parameter space in which the former branching ratio surpasses the latter (the small island of black in the yellow region near the left side of the plot for $\mr = 120$~GeV), despite the relative size of the couplings favoring decays to the latter.\footnote{The appearance of these islands depends somewhat on the grid spacing of our scan.}

\begin{figure}
\centering
	\includegraphics[width=0.3\textwidth]{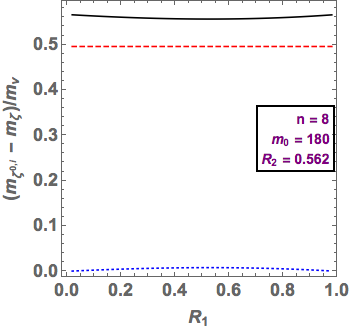}\hspace*{1em}
	\includegraphics[width=0.3\textwidth]{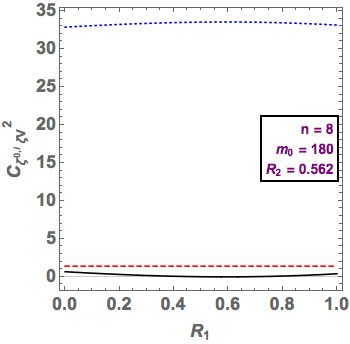}\hspace*{1em}
	\includegraphics[width=0.3\textwidth]{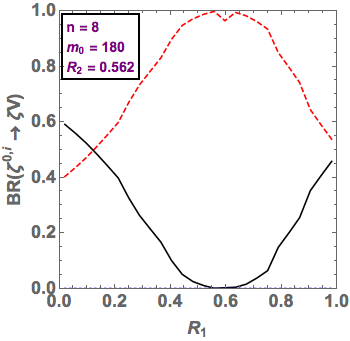}
	\caption{Scalar mass splittings normalized to the emitted gauge boson mass (left), coupling strengths squared (middle), and branching ratios (right) for $\zeta^{0,i}$ as a function of $R_1$ in the $n = 8$ model for $\mr = 180$~GeV and $R_2 = 0.562$ (compare Fig.~\ref{fig:decays_Z0I_n8}). Shown are the decays to $H_1^{\pm} W^{\mp}$ (solid black curves), $\zeta^{0,r} Z$ (dashed red curves), and $H_2^{\pm} W^{\mp}$ (dotted blue curves).  Note that the $\zeta^{0,i} H_1^{\pm} W^{\mp}$ squared coupling (solid black curve) touches zero near $R_1 \simeq 0.6$.}
	\label{fig:Z0I_BR_Dm_C_n8_m180}
\end{figure}

To illustrate the interplay of coupling strengths and mass splittings in the pattern of branching ratios, we examine in more detail two slices through the parameter space of Figs.~\ref{fig:decays_H22_n6} and \ref{fig:decays_Z0I_n8}.  In Fig.~\ref{fig:Z0I_BR_Dm_C_n8_m180} we show a slice through the $\mr = 180$~GeV panel of Fig.~\ref{fig:decays_Z0I_n8} (decays of $\zeta^{0,i}$ in the $n=8$ model) at $R_2 = 0.562$.  The relevant mass splittings (left plot in Fig.~\ref{fig:Z0I_BR_Dm_C_n8_m180}, normalized to the mass of the daughter gauge boson) are nearly constant with $R_1$, as are the $\zeta^{0,i} H_2^{\pm} W^{\mp}$ and $\zeta^{0,i} \zeta^{0,r} Z$ couplings (middle plot in Fig.~\ref{fig:Z0I_BR_Dm_C_n8_m180}).  Instead, the variation in the branching ratios is due to the variation in the $\zeta^{0,i} H_1^{\pm} W^{\mp}$ coupling, which passes through zero around $R_1 \simeq 0.6$.  This behavior can be traced back to the variation of the mixing angle $\alpha_1$.  This results in a big dip in the $\zeta^{0,i} \to H_1^{\pm} W^{\mp}$ branching ratio in favor of the decay $\zeta^{0,i} \to \zeta^{0,r} Z$ (right plot in Fig.~\ref{fig:Z0I_BR_Dm_C_n8_m180}).  For small $R_1$ where the decay to $H_1^{\pm} W^{\mp}$ does dominate over $\zeta^{0,r} Z$, the $\zeta^{0,i} H_1^{\pm} W^{\mp}$ coupling is still smaller than the  $\zeta^{0,i} \zeta^{0,r} Z$ coupling; the larger branching fraction is instead due to the larger available phase space for the former mode.
Note also that, although the $\zi H_2^{\pm} W^{\mp}$ coupling is significantly stronger than either of the others, the resulting decay is severely kinematically suppressed and never dominates.  This is a generic feature of our models: due to the pattern of mixings among the isospin eigenstates, the gauge couplings between pairs of scalars tend to be largest when their mass splittings are smallest.  The exception is the $\zeta^{0,i} \zeta^{0,r} Z$ coupling, which does not depend on any mixing angles.

\begin{figure}
\centering
	\includegraphics[width=0.3\textwidth]{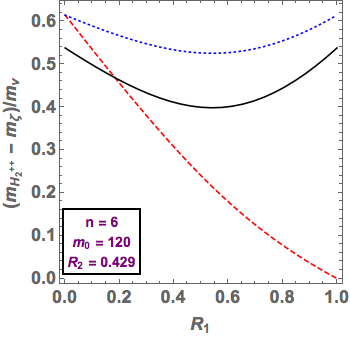}\hspace*{1em}
	\includegraphics[width=0.3\textwidth]{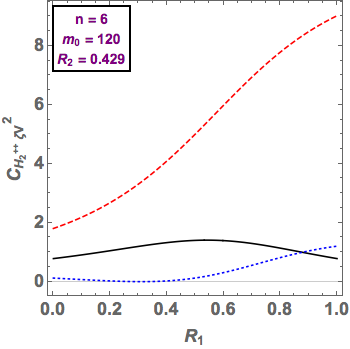}\hspace*{1em}
	\includegraphics[width=0.3\textwidth]{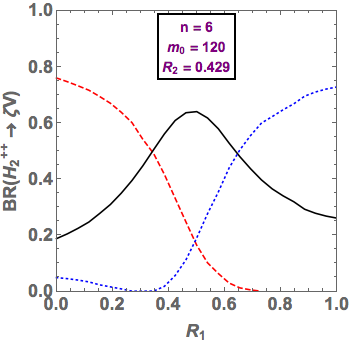}
	\caption{Scalar mass splittings normalized to the emitted gauge boson mass (left), coupling strengths squared (middle), and branching ratios (right) for $H_2^{++}$ as a function of $R_1$ in the $n = 6$ model for $\mr = 120$~GeV and $R_2 = 0.429$ (compare Fig.~\ref{fig:decays_H22_n6}).  Shown are the decays to $\zeta^{+3} W^-$ (dashed red curves), $H_1^{++} Z$ (solid black curves), and $H_1^+ W^+$ (dotted blue curves).  Note that the $H_2^{++} H_1^- W^-$ squared coupling (dotted blue curve) touches zero near $R_1 \simeq 0.3$.}
	\label{fig:H22_BR_Dm_C_n6_m120}
\end{figure}

In Fig.~\ref{fig:H22_BR_Dm_C_n6_m120} we show a slice through the $\mr = 120$~GeV panel of Fig.~\ref{fig:decays_H22_n6} (decays of $H_2^{++}$ in the $n = 6$ model) at $R_2 = 0.429$.  The three-band structure in Fig.~\ref{fig:decays_H22_n6} emerges in a natural way from the interplay between couplings and mass splittings.  At small $R_1$ the $H_2^{\pm\pm} \zeta^{\mp3} W^{\pm}$ coupling is sizable and the available phase space is large, so this decay dominates.  As $R_1$ increases, the mass of $\zeta^{+3}$ increases (see also Fig.~\ref{fig:massspectra}), and the $H_2^{\pm\pm} \to \zeta^{\pm3} W^{\mp}$ branching fraction drops due to the squeezing of the phase space, in spite of the growth of the $H_2^{\pm\pm} \zeta^{\mp3} W^{\pm}$ coupling.  Meanwhile, the $H_2^{\pm\pm} H_1^{\mp} W^{\mp}$ coupling is passing through zero around $R_1 \simeq 0.3$.  This allows the decay to $H_1^{\pm\pm} Z$ to dominate for $R_1$ between about 0.3 and 0.7.  For high values of $R_1$, the rising $H_2^{\pm\pm} H_1^{\mp} W^{\mp}$ coupling and its larger phase space allow the decay to $H_1^{\pm} W^{\pm}$ to become dominant.

Finally, we note that the maximum allowed mass splitting $\DmMAX \equiv \mi - \mr$ is considerably smaller in the $n=8$ model than in the $n=6$ model (see Table~\ref{tbl:Dm0MAX}).  While on-shell decays first appear in the $\mr = 120$~GeV plots for the $n=6$ model, they only appear in the $\mr = 300$~GeV plots for the $n=8$ model.

\section{Collider study}
\label{sec:collider}

LHC constraints on models with stable neutral particles (i.e., WIMP dark matter candidates) typically arise from searches for excess events with large missing transverse energy ($\met$) recoiling against some combination of visible matter. These types of signatures have been studied both in the context of supersymmetry and in generic mono-jet searches by both ATLAS and CMS using LHC8 data.  Searches for ${\rm jets}+\met$ are typically only constraining for signals produced via QCD-strength interactions because of the large $W/Z+{\rm jets}$ backgrounds; nevertheless, we include them here for completeness.  
Signatures involving ${\rm leptons}+\met$, for which the SM backgrounds are smaller, are typically more constraining for models in which the new physics is produced with electroweak-strength cross sections, as in our models.  These ${\rm leptons}+\met$ signatures will be responsible for the direct-search constraints on our models.
We take advantage of the large number of possible pairs of scalars that can be produced via $s$-channel $\gamma$, $Z$, or $W$ exchange in our models by simulating all such processes and combining the resulting signal events.

The following searches were reproduced:
\begin{itemize}\vspace{-0.15cm}
  \item ATLAS opposite-sign dileptons with $\met$ and no jets~\cite{TheATLAScollaboration:2013hha};
\vspace{-0.15cm}
  \item ATLAS trilepton plus $\met$~\cite{ATLAS:2013rla,Aad:2014nua} (the two studies employ distinct methodologies, and are reproduced separately);
\vspace{-0.15cm}
  \item ATLAS four or more leptons~\cite{ATLAS:2013qla};
\vspace{-0.15cm}
  \item ATLAS dileptons with razor variables~\cite{TheATLAScollaboration:2013via};
\vspace{-0.15cm}
  \item ATLAS hadronic di-$\tau$ plus $\met$~\cite{ATLAS:2013yla};
\vspace{-0.15cm}
  \item ATLAS same-sign dileptons plus jets~\cite{ATLAS:2013tma};
\vspace{-0.15cm}
  \item ATLAS monojet~\cite{ATLAS:2012zim,TheATLAScollaboration:2013aia};
\vspace{-0.15cm}
  \item ATLAS multi-jets plus $\met$~\cite{TheATLAScollaboration:2013fha};
\vspace{-0.15cm}
  \item CMS multi-leptons (dilepton, trilepton, multi-lepton) with $\met$~\cite{CMS:2013dea};
\vspace{-0.15cm}
  \item CMS monojet~\cite{CMS:rwa}.
\end{itemize}

We performed the simulation following the methodology of Ref.~\cite{Martin:2014qra}.
For each of the four mass slices, $\mr = 80$, 120, 180, and 300~GeV in each model, we generated parameter points on a grid in $R_1$ and $R_2$, focusing on the region that is allowed by the theoretical and indirect experimental constraints.  We also simulated a few additional points for added sensitivity in regions where the experimental sensitivity was varying rapidly. The parameter points that we simulated are shown as green dots on our exclusion plots.  

We do not simulate parameter points for which $\Delta m_0 \equiv \mi - \mr < 10$~GeV.  As we will see, this compressed-spectrum region is particularly difficult to detect because the leptons resulting from the decays tend to be too soft to pass the selection cuts.  Indeed, we will find that parameter regions with $\Delta m_0 \lesssim 20$~GeV cannot be excluded using the LHC8 searches that we recast.  This parameter region could be picked up by ${\rm jets}+\met$ searches, but we find that the scalar pair production cross sections are too small for LHC8 to have any sensitivity in this channel.  Very small mass splittings could result in macroscopic decay lengths for charged scalars; while dedicated searches for such signatures exist, we will not consider them here.

Using \texttt{MadGraph5\_aMC@NLO} and our \texttt{FeynRules} model files, we generated 50,000 events for each of the possible pairs of scalars that can be produced via $s$-channel $W^\pm$ and $\gamma^*/Z$ exchange, in association with zero or one hard jet(s).  This amounted to 32 possible scalar pairings in the $n=6$ model and 44 in the $n=8$ model.  Each simulated event sample was then matched and merged, decayed, showered, and hadronized using \texttt{Pythia 6}~\cite{Sjostrand:2006za}. The events were then passed through the \texttt{Delphes 3}~\cite{deFavereau:2013fsa} detector simulation multiple times: for each of the experimental searches that we recast, we adjusted the identification efficiencies and jet algorithm settings in \texttt{Delphes 3} to match the working point used in the experimental analysis. We then used the \texttt{Seer}~\cite{Martin:2015hra} analysis program to apply the trigger and kinematic cuts of each experimental analysis and determine the total signal cross section from all production processes that fall into each of the signal regions. Additional Gaussian smearing of the $\met$ beyond that already present in \texttt{Delphes 3} was introduced via \texttt{Seer} to account for the effect of pileup~\cite{ATL-PHYS-PUB-2013-004}.  This was necessary in order to reproduce the cut-flow tables for each of the experimental searches.  

At each simulated parameter point, we select as most sensitive the experimental search that yields the largest value of $\log_{10}(N_{\rm sig}/N_{95})$, where $N_{\rm sig}$ is the number of signal events computed from the cross section in the signal region for the appropriate integrated luminosity and $N_{95}$ is the corresponding 95\% confidence level exclusion threshold from the experimental analysis.  If $\log_{10}(N_{\rm sig}/N_{95}) > 0$ for the most sensitive search, the parameter point is excluded.  We do not combine significances from different search channels---to do so would require knowledge of the statistical correlations among the various searches which we do not have.  In between parameter points, we interpolate linearly in the value of $\log_{10}(N_{\rm sig}/N_{95})$ found for the most sensitive search at each point.

We find that the most sensitive analyses are the ATLAS dilepton, trilepton and four-lepton searches. Several other searches also exclude parameter space for low values of $\mr$, but this is primarily due to the large pair production rates when the scalar masses are low, which overcome the inherently smaller acceptances and/or larger backgrounds in these searches. 

Our results are shown in Fig.~\ref{fig:results_n6} for the $n = 6$ model and Fig.~\ref{fig:results_n8} for the $n=8$ model.  For each of the mass slices $\mr = 80$, 120, 180, and 300~GeV, we show the excluded region of the $R_1$--$R_2$ plane (darker-colored regions above and to the left of the solid black line).  The dashed black line indicates how the excluded region would expand if the signal cross sections were all increased by a $k$-factor of 1.2.  As before, the parameter region below the thick teal curve is allowed by the theoretical and indirect experimental constraints.  In each plot the thick horizontal light pink line indicates the $R_2$ value at which $\Delta m_0 \equiv \mi - \mr = 20$~GeV, below which the spectrum is highly compressed.

We now describe the results for each model.

\subsection{Results for the $n=6$ model}

In the $n=6$ model, the parameter space is strongly excluded at low $\mr$ so long as $\Delta m_0 > 20$~GeV.  As $\mr$ increases, the excluded region shrinks until only a small sliver of parameter space around $R_2 \sim 0.3$ is excluded for $\mr = 180$~GeV.  LHC8 makes no exclusion at $\mr = 300$~GeV or above.

\begin{figure}
\centering
	\begin{subfigure}[t]{0.44\textwidth}
	\includegraphics[width=\textwidth]{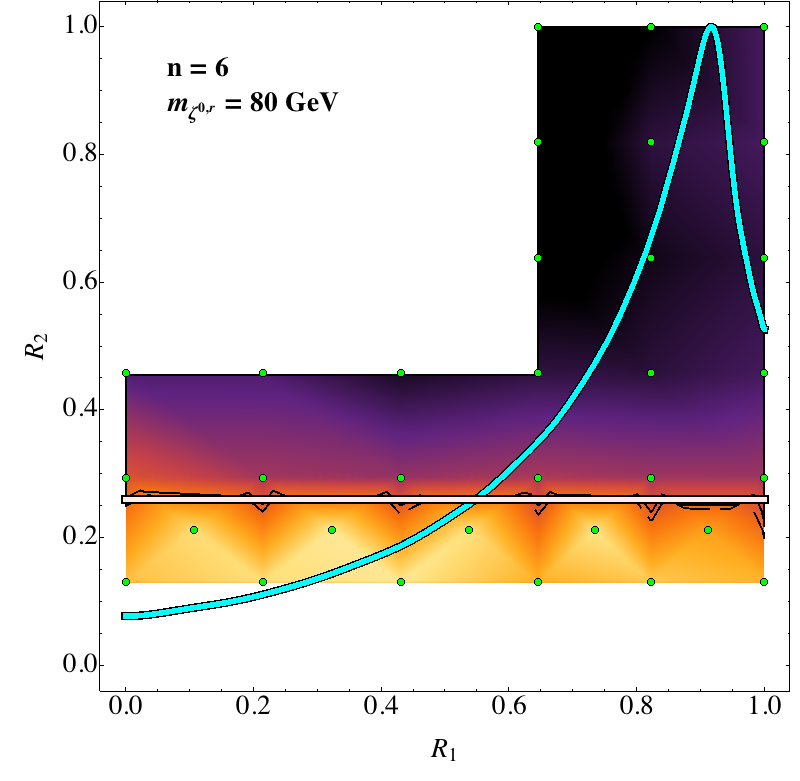}
	\end{subfigure}
	\begin{subfigure}[t]{0.44\textwidth}
	\includegraphics[width=\textwidth]{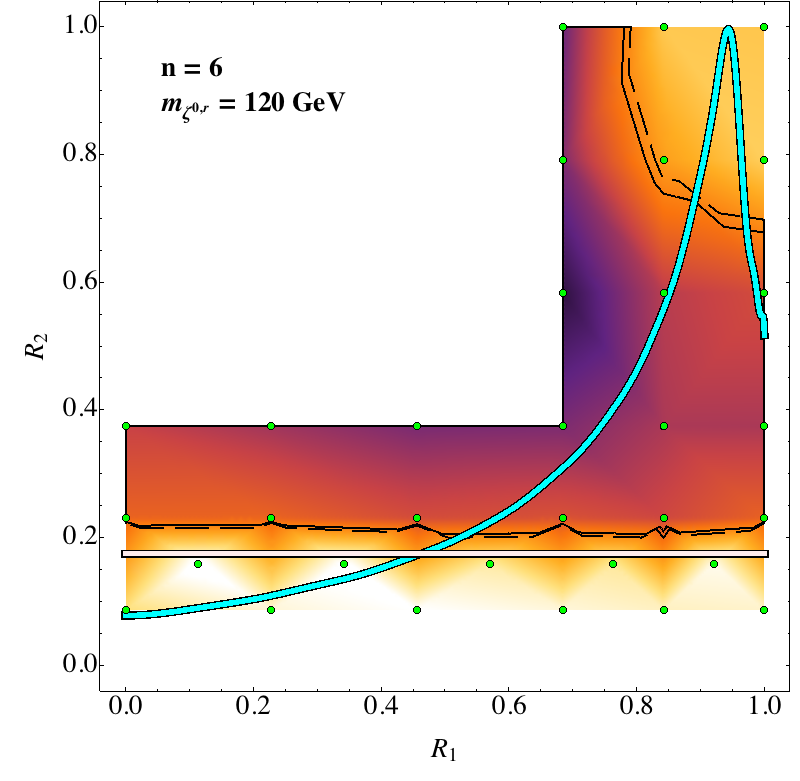}
	\end{subfigure}
	\begin{subfigure}[t]{0.1\textwidth}
	\includegraphics[width=\textwidth]{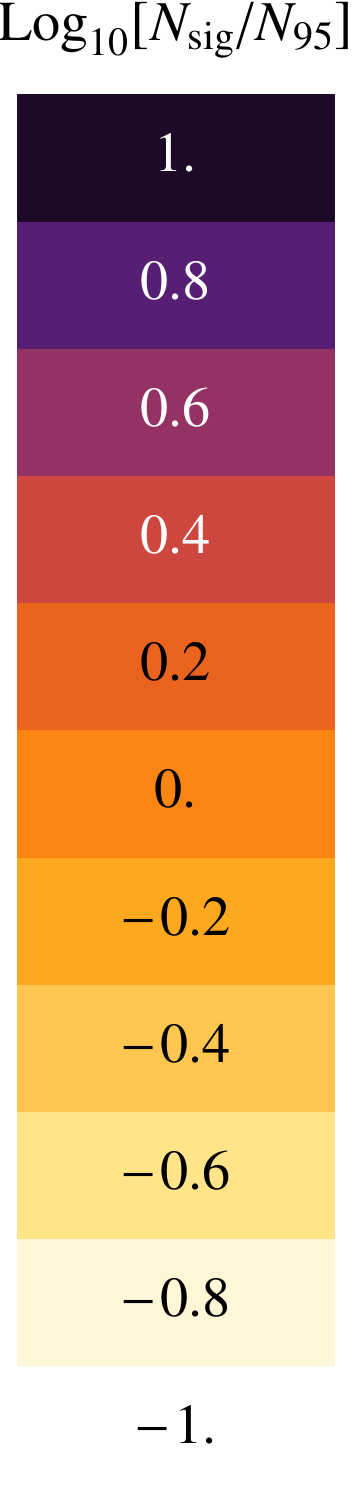}
	\end{subfigure}
	
	\begin{subfigure}[t]{0.44\textwidth}
	\includegraphics[width=\textwidth]{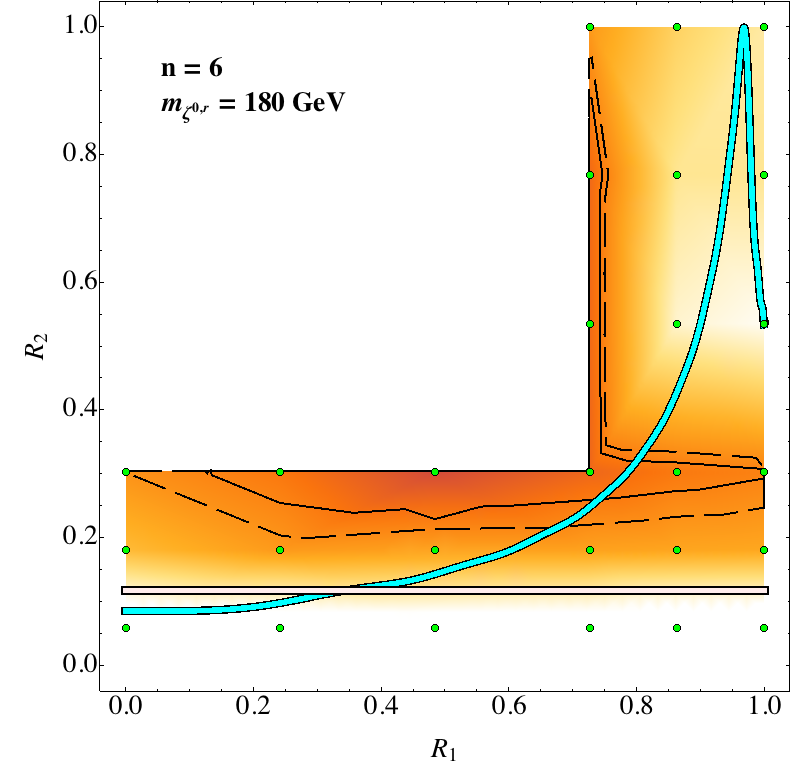}
	\end{subfigure}
	\begin{subfigure}[t]{0.44\textwidth}
	\includegraphics[width=\textwidth]{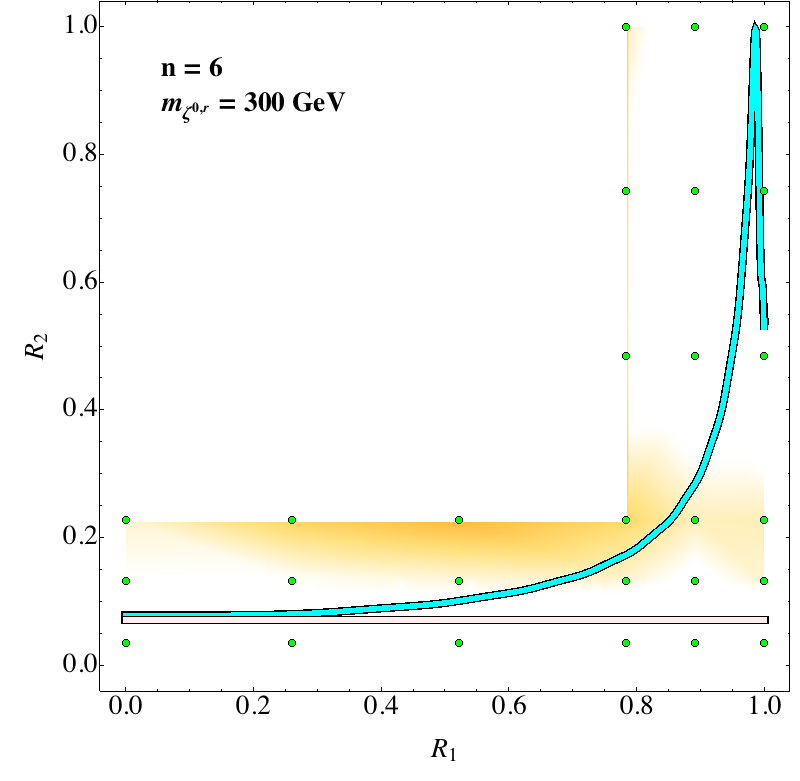}
	\end{subfigure}
	\begin{subfigure}[t]{0.1\textwidth}
	\includegraphics[width=\textwidth]{./Legend}
	\end{subfigure}
	\caption{Combined exclusion regions in the $n=6$ model from all examined LHC8 analyses. The darker-colored regions above and to the left of the solid black line are excluded at 95\% confidence level by at least one analysis.  The dashed black line indicates how the excluded region would expand if the signal cross sections were all increased by a $k$-factor of 1.2. The parameter region below the thick teal curve is allowed by the theoretical and indirect experimental constraints.  The thick horizontal light pink line indicates the $R_2$ value at which $\Delta m_0 \equiv \mi - \mr = 20$~GeV.}
	\label{fig:results_n6}
\end{figure}

For $\mr=80$~GeV, shown in the upper left panel of Fig.~\ref{fig:results_n6}, the total scalar pair production cross section before cuts ranges from $\sim30$~pb for small $R_2$ to $\sim16$~pb for large $R_2$.  The ATLAS ${\rm trilepton}+\met$ searches~\cite{ATLAS:2013rla,Aad:2014nua} provide the most sensitive exclusion for all parameter points scanned. Due to the large signal cross sections, even small acceptance rates result in large numbers of events in the experimental signal regions, resulting in the strong exclusion as indicated by dark colors.  The excluded region ends abruptly at $\Delta m_0 \simeq 20$~GeV.  For $R_2$ values below this boundary, the highest lepton $p_T$ is rarely above 10~GeV, unless the scalars are produced boosted or in association with an energetic jet, which is rare.  The ATLAS and CMS offline leptonic triggers are typically set at $p_T \geq 10$~GeV, resulting in a drastic plunge in acceptance.  Similarly, the $\met$ tends to be low when the spectrum is compressed; most (but not all) of the searches require $\met \geq 50$~GeV, further reducing sensitivity to the compressed-spectrum region.  This kinematic acceptance boundary is apparent in many of the exclusion plots for both the $n=6$ and $n=8$ models.

For $\mr=120$~GeV, shown in the upper right panel of Fig.~\ref{fig:results_n6}, the total scalar pair production cross section before cuts ranges from $\sim8.2$~pb for small $R_2$ to $\sim3.5$~pb for large $R_2$. The ATLAS ${\rm trilepton}+\met$ searches~\cite{ATLAS:2013rla,Aad:2014nua} provide the most sensitive exclusion for $R_2 \lesssim 0.5$, in particular the search region with same-flavor opposite-sign dilepton mass smaller than the $Z$ mass, which picks up the off-shell $Z$ boson decays. For $R_2 \gtrsim 0.5$ and $R_1 \gtrsim 0.75$, the most sensitive search is the ATLAS four-lepton search~\cite{ATLAS:2013qla}, due to the smaller backgrounds and the predominance of $\zeta^{0,i} \to \zeta^{0,r} Z$ and $H_2^+ \to H_1^+ Z$ decays in this region (see Figs.~\ref{fig:decays_Z0I_n6} and \ref{fig:decays_H21_n6}). The unexcluded region in the upper right corner of this plot results from the loss of the four-lepton signatures as the decays involving $W$ emission go on shell (the resulting two-lepton final states suffer from larger backgrounds).

For $\mr=180$~GeV, shown in the lower left panel of Fig.~\ref{fig:results_n6}, the total scalar pair production cross section before cuts ranges from $\sim1.7$~pb for small $R_2$ to $\sim0.7$~pb for large $R_2$.  The ATLAS ${\rm trilepton}+\met$ searches~\cite{ATLAS:2013rla,Aad:2014nua} are responsible for the exclusion between $R_2$ values of about 0.2 and 0.3, while the ATLAS four-lepton search~\cite{ATLAS:2013qla} is responsible for the thin strip of excluded parameter space near $R_1 \simeq 0.75$.  Within the parameter region allowed by theoretical and indirect experimental constraints, only a small sliver of parameter space around $R_2 \sim 0.3$ is excluded by the ATLAS ${\rm trilepton}+\met$ searches.  The largest contribution to the ${\rm trilepton}+\met$ signal is from the $H_2^{\pm\pm} H_2^\mp$ pair production mode, which decays in this region of parameter space mainly to $H_1^{\pm} W^{\pm} H_1^{\mp} Z$, followed by very soft decays of the $H_1^{\pm}$ states.  As $R_2$ increases from 0.3 to 0.4, the $H_2^{\mp}$ decay transitions to $\zeta^{0,r} W^{\mp}$ due to the nearby kinematic threshold (see Fig.~\ref{fig:decays_H21_n6}) so that the third lepton is lost.  The production cross sections for the heavier scalars also decrease with increasing $R_2$.

For $\mr = 300$~GeV, shown in the lower right panel of Fig.~\ref{fig:results_n6}, the total scalar pair production cross section before cuts ranges from $\sim0.2$~pb for small $R_2$ to $\sim0.08$~pb for large $R_2$. After accounting for branching ratios and acceptances, it is clear that there is simply insufficient luminosity collected at LHC8 to result in any sensitivity to the model for this and higher masses.

\subsection{Results for the $n=8$ model}

In the $n=8$ model, the parameter space at low $\mr$ is again excluded so long as $\Delta m_0 \gtrsim 20$~GeV.  The excluded region shrinks rapidly with increasing $\mr$ until LHC8 can make no exclusion at $\mr = 180$~GeV or above.  

\begin{figure}
\centering
	\begin{subfigure}[t]{0.44\textwidth}
	\includegraphics[width=\textwidth]{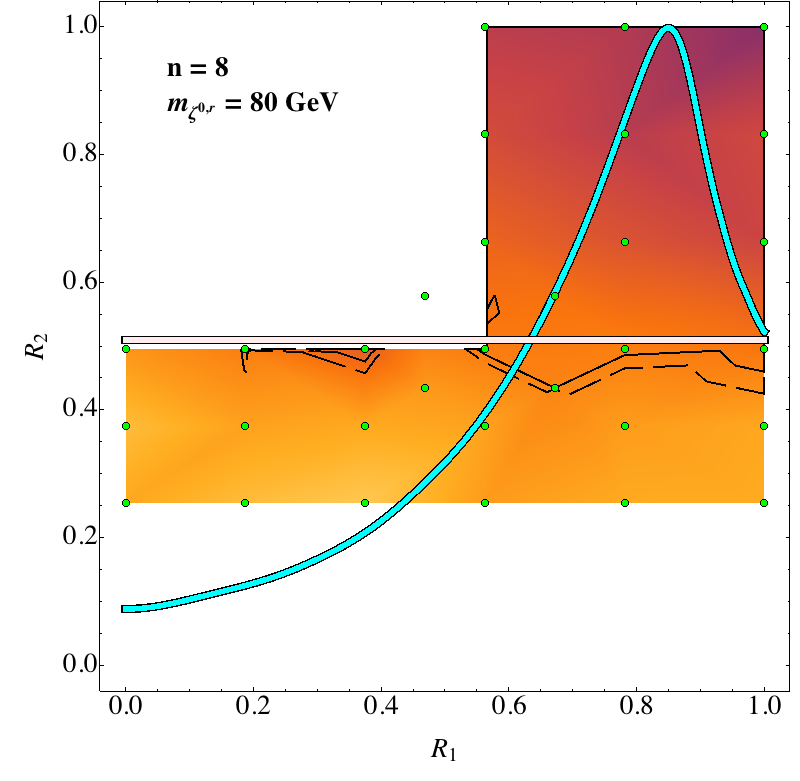}
	\end{subfigure}
	\begin{subfigure}[t]{0.44\textwidth}
	\includegraphics[width=\textwidth]{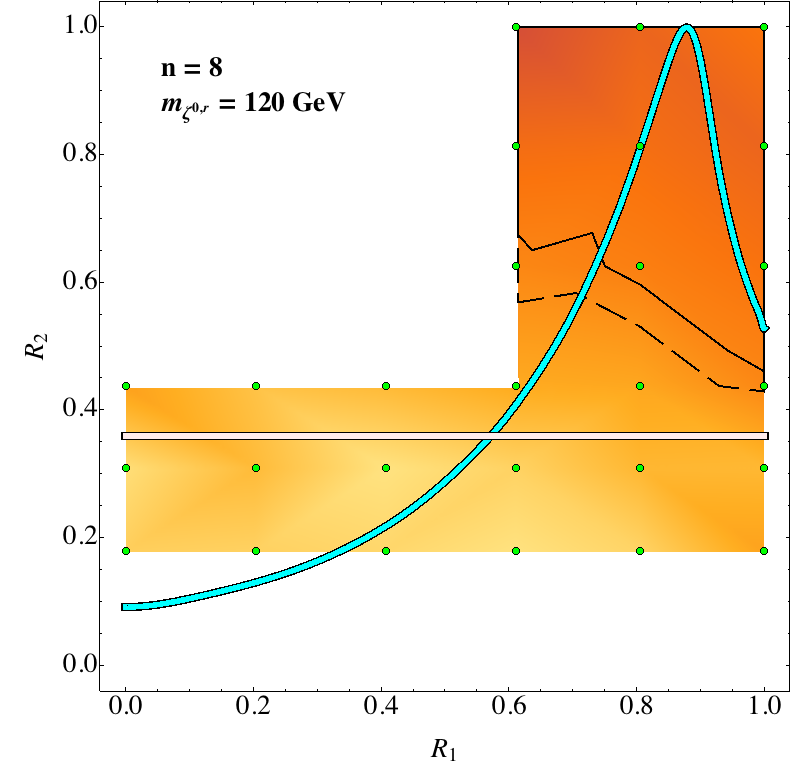}
	\end{subfigure}
	\begin{subfigure}[t]{0.1\textwidth}
	\includegraphics[width=\textwidth]{./Legend}
	\end{subfigure}
	
	\begin{subfigure}[t]{0.44\textwidth}
	\includegraphics[width=\textwidth]{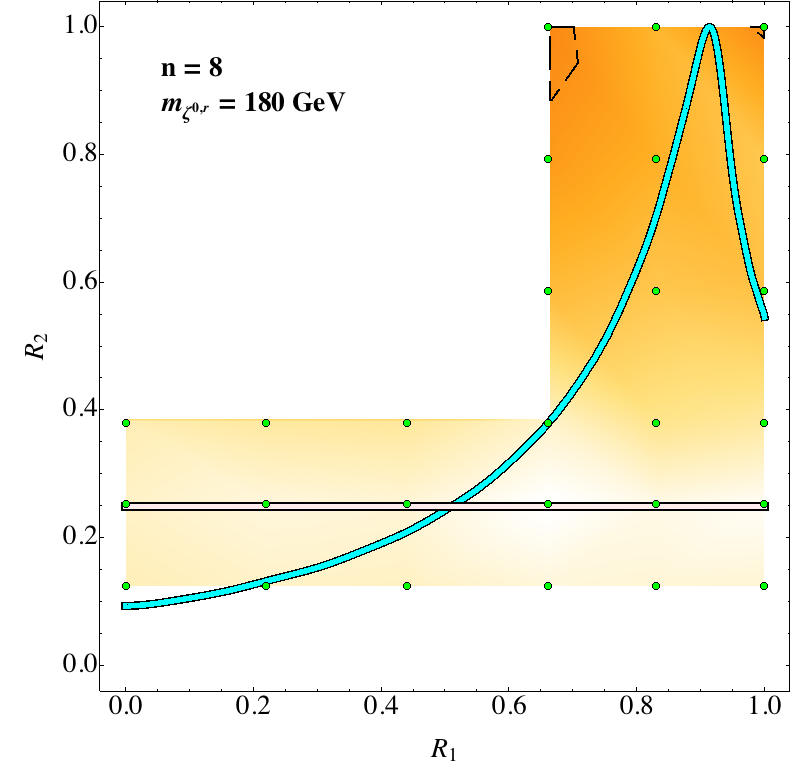}
	\end{subfigure}
	\begin{subfigure}[t]{0.44\textwidth}
	\includegraphics[width=\textwidth]{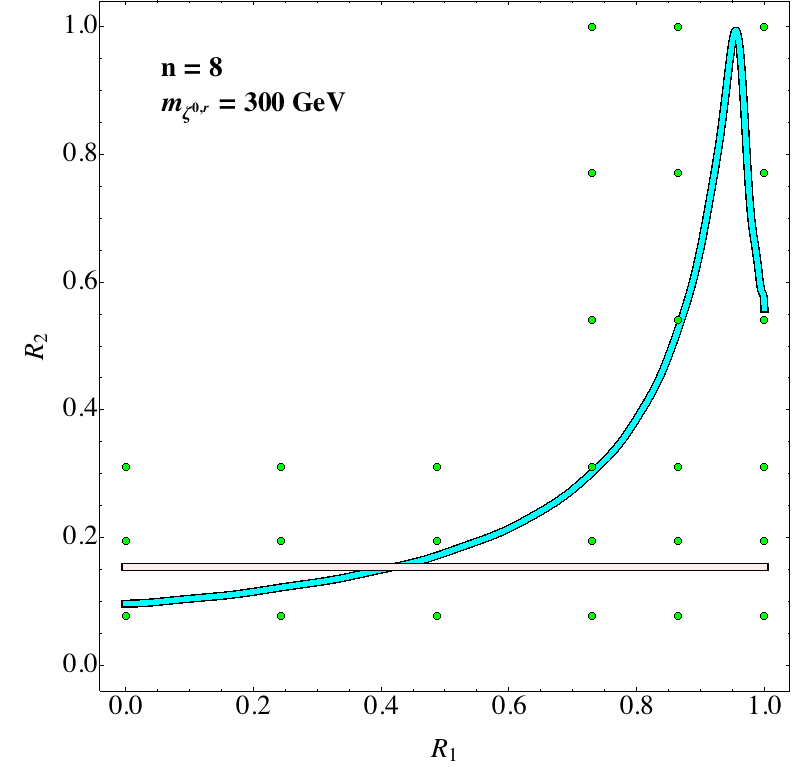}
	\end{subfigure}
	\begin{subfigure}[t]{0.1\textwidth}
	\includegraphics[width=\textwidth]{./Legend}
	\end{subfigure}
	\caption{Combined exclusion regions in the $n=8$ model from all examined LHC8 analyses. The darker-colored regions above the solid black line are excluded at 95\% confidence level by at least one analysis.  The dashed black line indicates how the excluded region would expand if the signal cross sections were all increased by a $k$-factor of 1.2. The parameter region below the thick teal curve is allowed by the theoretical and indirect experimental constraints.  The thick horizontal light pink line indicates the $R_2$ value at which $\Delta m_0 \equiv \mi - \mr = 20$~GeV.}
	\label{fig:results_n8}
\end{figure}

The most important difference between the two models is that the maximum mass splitting between the lightest and heaviest scalars, $\DmMAX \equiv \mi - \mr$, is considerably smaller in the $n=8$ model than in the $n=6$ model, by about a factor of two (see Table~\ref{tbl:Dm0MAX}).  Therefore, a much larger fraction of the allowed $n=8$ model parameter space lies in the compressed region with $\Delta m_0 < 20$~GeV (below the thick light pink horizontal line in Fig.~\ref{fig:results_n8}).  The smaller mass splittings over the entire $n=8$ parameter space result in softer leptons, jets, and $\met$ and hence overall lower acceptance rates in the LHC8 analyses.  This results in weaker exclusions in the $n=8$ model in spite of the larger number of scalars, their higher weak and electric charges (leading to more like-sign leptons from longer decay chains), and the correspondingly higher total scalar pair production cross sections before cuts.  The emitted leptons are just too soft.
 
For $\mr = 80$~GeV, shown in the upper left panel of Fig.~\ref{fig:results_n8}, the total scalar pair production cross section before cuts ranges from $\sim90$~pb for small $R_2$ to $\sim62$~pb for large $R_2$.  The entire parameter space with $R_2 \gtrsim 0.5$ is excluded.
Similarly to the $n=6$ model, the excluded region ends around $\Delta m_0 \simeq 20$~GeV, below which the highest lepton $p_T$ is rarely above 10~GeV, so that most events fail the offline leptonic triggers. The most sensitive exclusion comes from the published ATLAS ${\rm trilepton}+\met$ study~\cite{Aad:2014nua}. This study captures a broader range of signal characteristics than the corresponding conference note~\cite{ATLAS:2013rla}, and includes signal regions that are more sensitive to low dilepton invariant masses (called the $SR0\tau a$ bins 1--4 in Ref.~\cite{Aad:2014nua}). In particular, the best constraints on the $n=8$ model come from $SR0\tau a$ bin 2, in which the invariant mass of the same-flavor opposite-sign lepton pair lies between 12 and 40~GeV, the transverse mass of the remaining lepton and the $\met$ lies below 80~GeV, and $\met>90$~GeV. The most constraining bin in Ref.~\cite{ATLAS:2013rla} is similar, but with a softer cut on $\met$, which results in larger background. The ATLAS four-lepton search~\cite{ATLAS:2013qla} also redundantly excludes part of the parameter space, but loses sensitivity for larger values of $R_1$, where the dominant decay of $H_2^{\pm\pm}$ changes from $H_1^{\pm\pm} Z$ to $H_1^{\pm} W^{\pm}$.

For $\mr = 120$~GeV, shown in the upper right panel of Fig.~\ref{fig:results_n8}, the total scalar pair production cross section before cuts ranges from $\sim20$~pb for small $R_2$ to $\sim12$~pb for large $R_2$.  
The excluded region is understood similarly to the previous mass slice, with the same mechanisms at work.  When the smaller cross sections are combined with branching ratios and acceptance rates, the signal events in the published ATLAS ${\rm trilepton}+\met$ selection~\cite{Aad:2014nua} wind up falling into multiple signal bins, rather than being as concentrated in a single bin as in the $\mr = 80$~GeV slice. Since we do not have information on the correlations of the systematic uncertainties among different bins, we cannot employ the full usefulness of the $CL_s$ method to combine the bins to extend the exclusion boundaries. In addition, decays to pairs of off-shell $Z$ bosons are common, which reduces the trilepton signal from $WZ$ production. A four-lepton search is intrinsically less sensitive than a trilepton search due to the smaller leptonic branching ratio of the $Z$ compared to the $W$.

For $\mr = 180$~GeV, shown in the lower left panel of Fig.~\ref{fig:results_n8}, the total scalar pair production cross section before cuts ranges from $\sim4.7$~pb for small $R_2$ to $\sim2.7$~pb for large $R_2$.  None of the scanned parameter space is excluded by LHC8 searches.  A small region of parameter space at $R_2 \gtrsim 0.9$ and $R_1 \sim 0.7$ would be excluded if a signal $k$-factor of 1.2 were included. The sensitivity in this region comes from the ATLAS opposite-sign ${\rm dileptons}+\met$ search with no jets~\cite{TheATLAScollaboration:2013hha}, which is sensitive to the off-shell $Z$ decays from the $H_2^{\pm\pm}$ and $\zeta^{0,i}$ that are produced in association with $H_1$ states.  Jets from the decays of the $H_1$ states are soft enough that they evade the jet veto in the search.

For $\mr = 300$~GeV, shown in the lower right panel of Fig.~\ref{fig:results_n8}, the signal rate after cuts in any channel is less than 10\% of that required for an exclusion over the entire scanned parameter space.

\section{Discussion and Conclusions}\label{sec:conclusions}

We studied the direct-search constraints from LHC8 data on models in which the Higgs sector is extended by a $Z_2$-odd complex scalar electroweak multiplet $\multiplet$ with isospin $T=5/2$ ($n=6$) or $7/2$ ($n=8$) and the same hypercharge as the SM Higgs doublet. These models can be probed by recasting the dedicated searches for supersymmetric particles performed by the CMS and ATLAS collaborations. We showed that, even after imposing theoretical and indirect experimental constraints (mainly from the electroweak oblique parameters), the LHC8 searches provide nontrivial further constraints on the parameter space of the two models. 

The data from LHC8 excludes the majority of the remaining parameter space for $\mr = 80$--120~GeV in both the $n=6$ and $n=8$ models, except for parameter regions in which the entire spectrum is compressed within a mass splitting $\Delta m_0 < 20$~GeV.  The loss of sensitivity in the compressed spectrum region is caused by the extreme softness of the cascade decay products, leading most of the signal events to fail the trigger and minimum energy requirements for lepton tagging. The LHC8 constraints rapidly disappear for $\mr > 120$~GeV, until there is little sensitivity above $\mr = 180$~GeV. These results are in general agreement with the LHC8 constraints on the electroweak gauginos in supersymmetric models studied in Ref.~\cite{Martin:2014qra}, in which the same collection of LHC8 searches showed sensitivity to masses of the lightest supersymmetric particle primarily smaller than 100~GeV.

The LHC8 constraints on our models are much weaker than those on the scalar septet extension of the SM Higgs sector~\cite{Hisano:2013sn} studied in Ref.~\cite{Alvarado:2014jva}. Using similar searches from ATLAS and CMS, Ref.~\cite{Alvarado:2014jva} found that septet masses below $\sim400$~GeV were robustly excluded, and that multi-lepton searches provided the strongest constraints.  The dramatically enhanced LHC8 sensitivity to the septet model compared to our sextet and octet models is due to the nonzero vacuum expectation value carried by the septet and the mixing between the neutral scalar in the septet model and the SM Higgs boson in that model.  These features allow the lightest septet state to decay into SM particles, particularly into energetic gauge boson pairs, leading to final states with two to four high-energy leptons.  The mass reach of the LHC8 exclusion is then controlled by the decrease in the production cross section with increasing scalar masses.

With Run~2 of the LHC underway at 13~TeV center-of-mass energy, the next round of searches will likely provide stronger constraints on our models, extending to higher scalar masses. The pair production cross section for fixed scalar masses grows with increasing proton-proton center-of-mass energy due to the rise in the antiquark parton density. Higher integrated luminosity will also aid the searches, but with diminishing returns. The analyses that we use compare an observed event number to theoretical predictions for the background cross section; unless these backgrounds can be normalized from data, such searches suffer from systematic uncertainties that ultimately limit the benefit of greater luminosity. In addition, higher instantaneous luminosity will necessitate an increase in trigger thresholds, which will further reduce the sensitivity to the compressed spectrum regions.  Discovery prospects at a higher energy proton-proton collider would face similar issues.

Finally, we would be remiss not to mention the fact that the scalars in $\multiplet$ could be pair produced via $s$-channel photon and $Z$ exchange at a lepton collider such as the International Linear Collider (ILC)~\cite{Baer:2013cma}.  Such a machine would be able to probe scalar masses up to the pair production threshold $\sim\sqrt{s}/2$ with small backgrounds. ILC search prospects in the compressed spectrum region would require a dedicated study.

\begin{acknowledgments}
We would like to thank Mikhail Batygov, Wade Hong, Hugues Beauchesne, and Kevin Earl for their help in setting up and troubleshooting computing resources, and Katy Hartling for discussions about the models.
This work was supported by the Natural Sciences and Engineering Research Council of Canada.
\end{acknowledgments}

\appendix

\section{Masses and mixing angles}
\label{app:massesmixing}

In this section we give some of the mathematical details used in the derivation of the mass spectrum and mixing angles in Sec.~\ref{sec:models}.

For a complex scalar multiplet $\multiplet$ with hypercharge $Y=1$ (normalized so that $Q = T^3 + Y/2$), the most general gauge-invariant and $Z_2$-invariant renormalizable scalar potential was given in Eq.~(\ref{eq:conjugatepotential}), in which $\widetilde \Phi =  i \sigma^2 \Phi^*$ and $\widetilde \multiplet = C \multiplet^*$ are the conjugate multiplets.  Here $\sigma^2$ is the second Pauli matrix and the conjugation matrix $C$ for the large multiplet is an anti-diagonal $n \times n$ matrix. For $n=6$ and 8, the matrix $C$ is given by
\begin{equation}
C_{(n = 6)} = \left(\begin{matrix} 0 & 0 & 0 & 0 & 0 & 1\\
0 & 0 & 0 & 0 & -1 & 0\\
0 & 0 & 0 & 1 & 0 & 0\\
0 & 0 & -1 & 0 & 0 & 0\\
0 & 1 & 0 & 0 & 0 & 0\\
-1 & 0 & 0 & 0 & 0 & 0
\end{matrix}\right),  \qquad
C_{(n = 8)} = \left(\begin{matrix}0 & 0 & 0 & 0 & 0 & 0 & 0 & 1\\
0 & 0 & 0 & 0 & 0 & 0 & -1 & 0\\
0 & 0 & 0 & 0 & 0 & 1 & 0 & 0\\
0 & 0 & 0 & 0 & -1 & 0 & 0 & 0\\
0 & 0 & 0 & 1 & 0 & 0 & 0 & 0\\
0 & 0 & -1 & 0 & 0 & 0 & 0 & 0\\
0 & 1 & 0 & 0 & 0 & 0 & 0 & 0\\
-1 & 0 & 0 & 0 & 0 & 0 & 0 & 0
\end{matrix}\right).
\label{appeq:Cmatrix}
\end{equation}

Taking $\lambda_4$ real and working in unitarity gauge, the term involving $\lambda_4$ in the scalar potential of Eq.~(\ref{eq:conjugatepotential}) reduces to
\begin{equation}
	\lambda_4\, \widetilde{\Phi}^\dag \tau^a \Phi\, \multiplet^\dag T^a \widetilde{\multiplet} + \mathrm{h.c.} 
	= \frac{1}{4}\lambda_4 (h + v)^2\left[\multiplet^\dag T^{-} \widetilde{\multiplet} 
	+ \widetilde{\multiplet}^\dag T^{+} \multiplet\right],
\end{equation}
where $T^{\pm} = T^1 \pm i T^2$. For $n=6$ the generators $T^a$ are given by
\begin{equation}
T_{(n = 6)}^{+} = \left(\begin{matrix}0 & \sqrt{5} & 0 & 0 & 0 & 0 \\
0 & 0 & 2 \sqrt{2} & 0 & 0 & 0 \\
0 & 0 & 0 & 3 & 0 & 0 \\
0 & 0 & 0 & 0 & 2 \sqrt{2} & 0 \\
0 & 0 & 0 & 0 & 0 & \sqrt{5} \\
0 & 0 & 0 & 0 & 0 & 0\end{matrix}\right) = \left(T_{(n = 6)}^{-}\right)^{\dagger},
\end{equation}
\begin{equation}
T_{(n = 6)}^3 = \frac{1}{2}\,\mathrm{diag}\left(5,\,3,\,1,\,-1,\,-3,\,-5\right),
\end{equation}
while for $n=8$ they are
\begin{equation}
T_{(n = 8)}^{+} = \left(\begin{matrix}0 & \sqrt{7} & 0 & 0 & 0 & 0 & 0 & 0 \\
0 & 0 & 2 \sqrt{3} & 0 & 0 & 0 & 0 & 0 \\
0 & 0 & 0 & \sqrt{15} & 0 & 0 & 0 & 0 \\
0 & 0 & 0 & 0 & 4 & 0 & 0 & 0 \\
0 & 0 & 0 & 0 & 0 & \sqrt{15} & 0 & 0 \\
0 & 0 & 0 & 0 & 0 & 0 & 2 \sqrt{3} & 0 \\
0 & 0 & 0 & 0 & 0 & 0 & 0 & \sqrt{7} \\
0 & 0 & 0 & 0 & 0 & 0 & 0 & 0\end{matrix}\right) = \left(T_{(n = 8)}^{-}\right)^{\dagger},
\end{equation}
\begin{equation}
T_{(n = 8)}^3 = \frac{1}{2}\,\mathrm{diag}\left(7,\,5,\,3,\,1,\,-1,\,-3,\,-5,\,-7\right).
\end{equation}

The terms $\multiplet^\dag T^{-} \widetilde{\multiplet}$ and $\widetilde{\multiplet}^\dag T^{+} \multiplet$ split the masses of $\zeta^{0,r}$ and $\zeta^{0,i}$ and cause mixing between states with the same electric charge but different isospin.
For $n=6$ or 8, these two terms can be written as
\begin{eqnarray}
	\multiplet^\dag T^{-} \widetilde{\multiplet} &=& \frac{n}{2}(-1)^{n/2+1}\zeta^{0*}\zeta^{0*} + \sum_{Q=1}^{n/2} \sqrt{n^2-4Q^2} (-1)^{n/2+Q+1}\,\zeta^{+Q*}\zeta^{-Q*}, \nonumber \\
	\widetilde{\multiplet}^\dag T^{+} \multiplet &=& \frac{n}{2}(-1)^{n/2+1}\zeta^{0}\zeta^{0} + \sum_{Q=1}^{n/2} \sqrt{n^2-4Q^2} (-1)^{n/2+Q+1}\,\zeta^{+Q}\zeta^{-Q}.
\end{eqnarray}
Writing the neutral state $\zeta^0$ in terms of its real and imaginary components, $\zeta^0 = (\zeta^{0,r} + i\zeta^{0,i})/\sqrt{2}$, we find a mass splitting between the components,
\begin{eqnarray}
	m_{\zeta^{0,r}}^2 &=& M^2 + \frac{1}{2}v^2\left[\lambda_2 + \frac{1}{4}\lambda_3 + \frac{n}{2}(-1)^{n/2+1}\lambda_4\right] \equiv M^2 + \frac{1}{2}v^2\Lambda_n, \nonumber \\
	m_{\zeta^{0,i}}^2 &=& M^2 + \frac{1}{2}v^2\left[\lambda_2 + \frac{1}{4}\lambda_3 + \frac{n}{2}(-1)^{n/2}\lambda_4\right] = m_{\zeta^{0,r}}^2 + \frac{n}{2}(-1)^{n/2}v^2\lambda_4.
\end{eqnarray}

The mass matrices for the pairs of scalars with positive electric charge $Q=1, \ldots, \frac{n}{2}-1$ are given in the basis $(\zeta^{+Q}, \zeta^{-Q*})$ by
\begin{equation}
	M_Q^2 
	= \left(\begin{matrix}M^2 + \frac{1}{8}v^2(4\lambda_2 - (2Q-1)\lambda_3) & \frac{1}{4}v^2\lambda_4\sqrt{n^2-4Q^2}\,(-1)^{n/2+Q+1}\\ \frac{1}{4} v^2\lambda_4\sqrt{n^2-4Q^2}\,(-1)^{n/2+Q+1}  & M^2 + \frac{1}{8}v^2(4\lambda_2 + (2Q+1)\lambda_3)\end{matrix}\right),
\end{equation}
which we diagonalize to find the mass eigenvalues,
\begin{eqnarray}
	m_{H_{1,2}^{Q}}^2 &=& M^2 + \frac{1}{2}v^2\left(\lambda_2 + \frac{1}{4}\lambda_3 \mp \frac{1}{2}\sqrt{Q^2\lambda_3^2 + (n^2-4Q^2)\lambda_4^2}\right) \nonumber \\
	&=& m_{\zeta^{0,r}}^2 + \frac{1}{4}v^2\left(n(-1)^{n/2}\lambda_4 \mp \sqrt{Q^2\lambda_3^2 + (n^2-4Q^2)\lambda_4^2}\right).
\end{eqnarray}
The mass eigenstates $H_1^{Q}$ and $H_2^{Q}$ are defined in terms of the weak eigenstates by Eq.~(\ref{eq:alphadef}) such that $H_1^{Q}$ is the lighter state and $H_2^{Q}$ is the heavier state.
The mixing angle $\alpha_Q \in\nobreak [-\frac{\pi}{2},\frac{\pi}{2}]$ is given by
\begin{eqnarray}
	\tan \alpha_Q &=& (-1)^{n/2+Q+1}\frac{Q\lambda_3 - \sqrt{Q^2\lambda_3^2 + (n^2-4Q^2)\lambda_4^2}}{\sqrt{n^2-4Q^2}\,\lambda_4} \nonumber \\
	&=& (-1)^{n/2+Q}\frac{\sqrt{n^2-4Q^2}\,\lambda_4}{Q\lambda_3 + \sqrt{Q^2\lambda_3^2 + (n^2-4Q^2) \lambda_4^2}}.
\end{eqnarray}

There is only one state with $Q=n/2$.  Its mass is given by
\begin{equation}
	m_{\zeta^{n/2}}^2 = M^2 + \frac{1}{8}v^2\left(4\lambda_2 - (2Q-1)\lambda_3\right) 
	= m_{\zeta^{0,r}}^2 - \frac{n}{8}v^2\left(\lambda_3 + 2 (-1)^{n/2+1}\lambda_4\right).
\end{equation}

\section{Details of the re-parameterization}
\label{app:reparam}

For this study, it is convenient to describe the parameter space in terms of the mass differences $\Delta m_0 \equiv \mr - \mi$ and $\Delta m_{\frac{n}{2}} \equiv m_{\zeta^{+\frac{n}{2}}} - \mr$, since these two mass splittings are monotonic in $-\lambda_3$ and $|\lambda_4|$. 
In terms of the original Lagrangian parameters, the mass splittings are given by
\begin{align}
	\Delta m_0 &\equiv \mi - \mr = \sqrt{\mr^2 + \frac{n}{2}(-1)^{\frac{n}{2}}v^2\lambda_4} - \mr\;,\label{appeq:Dm0}\\
	\Delta m_{\frac{n}{2}} & \equiv m_{\zeta^{+\frac{n}{2}}} - \mr = \sqrt{\mr^2 - \frac{n}{8}v^2\left[\lambda_3 + 2 (-1)^{\frac{n}{2}+1}\lambda_4\right]} - \mr\;.\label{appeq:Dmn20}
\end{align}
We defined normalized versions of these splittings, each lying in the range $[0,\,1]$, as
\begin{align}
	R_1 &\equiv \frac{\Delta m_{\frac{n}{2}}}{\Delta m_{0}} = \frac{m_{\zeta^{+\frac{n}{2}}} - \mr}{\mi - \mr}\;,&
	R_2 &\equiv \frac{\Delta m_{0}}{\DmMAX} = \frac{\mi - \mr}{(\mi - \mr)^{\mathrm{MAX}}}\;.
	\label{appeq:R1R2def}
\end{align}

The maximum value of $\Delta m_0$ allowed by the theoretical and indirect experimental constraints (mainly from the electroweak oblique parameters), $\DmMAX$, can be written parametrically for the $n=6$ and $n=8$ models as
\begin{align}
	\Delta m_{0}^{\mathrm{MAX} (n=6)} &= \begin{cases}2.2 + 0.94 \mr & \mr \leq 530\unit{GeV,}\\
	809.4 - 0.72 \mr+ 0.00024 \mr^2 & \mr > 530\unit{GeV,}\end{cases}\\
	\Delta m_{0}^{\mathrm{MAX} (n=8)} &= \begin{cases}6.37 + 0.41 \mr & \mr \leq 809\unit{GeV,}\\
	687.7 - 0.57 \mr+ 0.00017 \mr^2 & \mr > 809\unit{GeV.}\end{cases}
\end{align}
The numerical values of $\Delta m_{0}^{\mathrm{MAX}}$ for our chosen mass slices were given in Table~\ref{tbl:Dm0MAX}. 

The relations in Eqs.~(\ref{appeq:R1R2def}), (\ref{appeq:Dm0}), and (\ref{appeq:Dmn20}) can be inverted to obtain $\lambda_3$ and $\lambda_4$ as follows:
\begin{align}
	\label{appeq:l3reparam}\lambda_3 &= \frac{4 R_2\,\DmMAX}{n\,v^2}\left[2\left(1 - 2 R_1\right)\mr + \left(1 - 2 R_1^2\right)R_2\,\Delta m_{0}^{\mathrm{MAX}}\right], \\
	\label{appeq:l4reparam}\lambda_4 &= (-1)^{\frac{n}{2}} \frac{2 R_2\, \DmMAX}{n\,v^2}\left[2 \mr + R_2\, \DmMAX\right]\;.
\end{align}
The physical masses can also be expressed as
\begin{align}
	\nonumber \mi &= \mr + R_2 \Delta m_{0}^{\mathrm{MAX}}, \\
	\nonumber m_{H_{1,2}^{+Q}} &= \sqrt{\mr^2 + \frac{1}{2} R_2^2 (\Delta m_{0}^{\mathrm{MAX}})^2 + R_2\left[\mr \mp \frac{\mathbb{M}}{2 n}\right] \Delta m_{0}^{\mathrm{MAX}}}, \\
	\label{appeq:newmasses} m_{\zeta^{+\frac{n}{2}}} &= \mr + R_1 R_2 \Delta m_{0}^{\mathrm{MAX}}\;,
\end{align}
where
\begin{displaymath}
	\mathbb{M} = \sqrt{\left(n^2 - 4 Q^2\right) \left(2 \mr + R_2 \Delta m_{0}^{\mathrm{MAX}}\right)^2 + 4 Q^2 \left[2 \left(1 - 2 R_1\right) \mr + \left(1 - 2 R_1^2\right) R_2 \Delta m_{0}^{\mathrm{MAX}}\right]^2}\;.
\end{displaymath}


\section{Compressed spectra versus \texttt{Pythia}}
\label{app:pythia_meson}

Our models include parameter regions in which the spectrum is highly compressed. In such regions, simulation problems can arise because the automatic parton-level computation of partial widths and branching ratios using \texttt{MadWidth}~\cite{Alwall:2014bza} does not take into account hadronization effects arising from decays involving a highly off-shell $W$ or $Z$ boson.

For example, consider the case of $H_1^\pm \rightarrow \zeta^{0,r} + W^\pm$ decay for a parameter point at which the mass splitting $m_{H_1^+} - \mr$ is larger than the mass of a $\pi^\pm$ but smaller than the mass difference of the $\pi^\pm - \pi^0$ system, as in Table~\ref{tab:spectra}.  Physically, a hadronically-decaying off-shell $W$ will emerge in this case as a single $\pi^{\pm}$.  However, \texttt{MadWidth} treats all decay products as bare partons and will decay the off-shell $W$ to a quark-antiquark pair. When \texttt{Pythia}~\cite{Sjostrand:2006za} hadronizes the resulting quark-antiquark pair, it tries to create two pions, finds that this is forbidden by the available phase space, and discards the event.

\begin{table}
\begin{center}
\begin{tabular}{C{6em}C{7em}C{8em}}
\hline \hline
State 			& Mass [GeV]	& $\Delta m_{i,i-1}$~[GeV]\\
\hline
$\zeta^{0,r}$	& 80.0000		& --\\
$H_1^\pm$ 		& 80.1960		& 0.196\\
$H_1^{\pm\pm}$ 	& 80.8320		& 0.636\\
$\zeta^{\pm 3}$ & 82.1524		& 1.3204\\
$H_2^{\pm\pm}$ 	& 89.2535		& 7.1011\\
$H_1^\pm	$	& 89.8254		& 0.5719\\
$\zeta^{0,i}$	& 90.0000		& 0.1746\\
\hline \hline
\end{tabular}
\caption{Mass spectrum and splittings for a sample point in the compressed-spectrum region, $\Delta m_0 = 10$~GeV. For this point, $n=6$, $\mr = 80$~GeV, $R_1 = 0.215$ and $R_2 = 0.130$.  Compare the charged pion mass at 0.140~GeV.}\label{tab:spectra}
\end{center}
\end{table}

This is a problem because it removes generated events that could otherwise potentially be detected via, e.g., a hard initial-state jet radiation.  We solve this problem by manually removing any hadronic $W$ branching ratio from the \texttt{MadGraph} \texttt{param\textunderscore{}card} files in which the $W$ invariant mass is less than the mass of two correspondingly-flavored mesons and rescaling the remaining $W$ decay branching ratios to sum to one. 

Should we worry about changing the potential detector signature of the very soft $W$ boson?  A simple kinematical argument indicates no. In the rest frame of the decaying scalar, the momentum of the outgoing decay products must satisfy, e.g., $|p_{\zeta^{0,r}}|+|p_{\rm jets}| \leq m_{H_1^{+}} - \mr$ for the above example. In this frame the decay product(s) of the $W$ will have momentum less than the scalar mass splitting. To make the $W$ decay product(s) visible in the detector, they must be boosted to an ultrarelativistic velocity; for example, a 10~GeV charged pion requires a relativistic boost factor $\gamma \simeq 70$.  This requires that the parent scalar itself be boosted by the same relativistic boost factor: for a scalar mass around 80~GeV, this implies a transverse momentum of more than 5~TeV, which is not even kinematically possible at the 8~TeV LHC.


\end{document}